\documentclass[floatfix,showpacs,amsmath,amsfonts,amssymb,aps,twocolumn,prb,superscriptaddress,nobibnotes]{revtex4-1}

\usepackage[table,dvipsnames]{xcolor}
\usepackage{graphicx}
\usepackage{physics}
\usepackage{braket}
\usepackage{mathtools}
\usepackage{diagbox}
\usepackage{array, multirow}
\usepackage[caption=false]{subfig}
\usepackage{dsfont}
\usepackage{bbm}
\usepackage{tikz}

\definecolor{darkred}{rgb}{0.6,0.,0.}
\definecolor{darkgreen}{rgb}{0.,0.5,0.}
\definecolor{darkblue}{rgb}{0.,0.,0.6}
\usepackage[breaklinks, colorlinks=true, linkcolor=darkred, citecolor=darkgreen, urlcolor=darkblue]{hyperref}

\begin{document}

\title{Localization renormalization and quantum Hall systems}
\author{Bartholomew Andrews}
\thanks{Contributed equally to this work.}
\affiliation{Department of Physics and Astronomy, University of California at Los Angeles, 475 Portola Plaza, Los Angeles, California 90095, USA}
\affiliation{Department of Physics, University of California at Berkeley, 100 South Dr, Berkeley, California 94720, USA}
\author{Dominic Reiss}
\thanks{Contributed equally to this work.}
\affiliation{Department of Physics and Astronomy, University of California at Los Angeles, 475 Portola Plaza, Los Angeles, California 90095, USA}
\author{Fenner Harper}
\affiliation{Department of Physics and Astronomy, University of California at Los Angeles, 475 Portola Plaza, Los Angeles, California 90095, USA}
\author{Rahul Roy}
\email{rroy@physics.ucla.edu}
\affiliation{Department of Physics and Astronomy, University of California at Los Angeles, 475 Portola Plaza, Los Angeles, California 90095, USA}
\date{\today}

\begin{abstract}
The obstruction to constructing localized degrees of freedom is a signature of several interesting condensed matter phases. We introduce a localization renormalization procedure that harnesses this property, and apply our method to distinguish between topological and trivial phases in quantum Hall and Chern insulators. By iteratively removing a fraction of maximally-localized orthogonal basis states, we find that the localization length in the residual Hilbert space exhibits a power-law divergence as the fraction of remaining states approaches zero, with an exponent of $\nu=0.5$. In sharp contrast, the localization length converges to a system-size-independent constant in the trivial phase. We verify this scaling using a variety of algorithms to truncate the Hilbert space, and show that it corresponds to a statistically self-similar expansion of the real-space projector. This result accords with a renormalization group picture and motivates the use of localization renormalization as a versatile numerical diagnostic for quantum Hall systems.
\end{abstract}

\maketitle
%\tableofcontents

%%%%%%%%%%%%%%%%%%%%%%%%
\section{Introduction} %
\label{sec:intro}      %
%%%%%%%%%%%%%%%%%%%%%%%%

% motivation
Renormalization group (RG) approaches are widely employed to distill the essential information from complex configurations and are an invaluable tool for studying the universal properties of systems close to criticality~\cite{Swingle16, White92, Wetterich93, Vidal08, Koch18}. Generally, when scaling relations indicate that a correlation length is the only relevant length scale close to a phase transition, we can leverage the statistical self-similarity of fluctuations up to this correlation scale, by gradually eliminating correlated degrees of freedom at all microscopic lengths. This is the basis of Kadanoff-Wilson RG, which is ubiquitous across a diverse body of research~\cite{Lihong21}. In recent years, a specific class of phase transitions, distinguished by the divergence or saturation of a characteristic localization length, which we call ``localization transitions", has attracted renewed interest. Examples include transitions arising from the topology-localization dichotomy, such as the plateau transition~\cite{Sbierski21, huang21, zhu19, zirnbauer19}, and those related to exotic localization phenomena, such as the thermal-MBL crossover~\cite{Garcia22, Thiery18, Guo18, Modak18}. In these systems, it is natural to construct a renormalization procedure based explicitly on the localization length to describe the critical phenomena. In analogy to traditional RG, when scaling relations indicate that the localization length is the pertinent length scale close to a phase transition, we can renormalize the system by gradually eliminating \emph{localized} degrees of freedom, which has analogous implications on the statistical self-similarity of the localized removal basis. In light of current research~\cite{Garcia22, Pu22, Li22, Bauer22, Liu23, Mildner23}, there is motivation to leverage the localization properties of critical systems in numerical methods, and apply such a procedure to study the growing array of localization transitions.

% summary
In this paper, we introduce a localization renormalization formalism and apply it to a selection of single-particle examples. Specifically, we focus on phase transitions resulting from the topology-localization dichotomy in quantum Hall systems. By iteratively removing the maximally-localized states from distinct sites in the system, we can induce a localization transition with a scaling given by the power-law divergence $\lim_{\rho\to 0, L\to\infty}\xi(\rho)\sim \rho^{-0.5}$, where $\xi$ is the localization length in the residual Hilbert space, $\rho$ is the fraction of states remaining, and $L$ is the linear system size. In contrast, the localization length converges to a system-size-independent constant in the trivial phase. We show that this scaling holds irrespective of the model used to describe the system, the metric used to quantify the localization length, and the way in which states are removed. Furthermore, we examine the expansion of the projector corresponding to the elimination of an orthogonal subset of maximally-localized states. Here, we reveal a direct correspondence between the power-law divergence of the localization length and a statistically self-similar expansion of the projector in real space. These results accord with an RG picture, where the effective length scale per state is increased by a factor of $\rho^{-1/2}$ on each step, and topological phases are identified by the presence of a phase transition. Apart from shedding additional light on the properties of integer quantum Hall systems, localization renormalization may be utilized as a numerical tool to characterize a wide variety of localization-delocalization transitions, including those in topologically trivial systems.

% structure
The structure of this paper is as follows. In Sec.~\ref{sec:loc_norm}, we define the localization renormalization procedure, and in Sec.~\ref{sec:qh_ex}, we apply it to single-particle case studies, representing continuous and discrete quantum Hall systems. In Sec.~\ref{sec:disc}, we then discuss the interpretations and scope of our results and finally, in Sec.~\ref{sec:conc}, we summarize the conclusions and outlook.

%%%%%%%%%%%%%%%%%%%%%%%%%%%%%%%%%%%%%%%%
\section{Localization Renormalization} %
\label{sec:loc_norm}                   %
%%%%%%%%%%%%%%%%%%%%%%%%%%%%%%%%%%%%%%%%

% system
We consider a $d$-dimensional quantum system, with single-particle Hilbert space $\mathcal{H}$, occupying a real-space region $A\in\mathbb{R}^d$ of linear extent $L$. To simply illustrate the steps, we focus on single-component single-particle orbitals residing in an isolated band.

% operators
We start by identifying a complete basis of wavefunctions for a given band $\{\ket{\psi}\}$. These single-particle orbitals are confined to regions surrounding points in space or sites on a lattice, and are maximally-localized according to a real-space metric --- conventionally, the distance-squared metric~\cite{marzari97}. We emphasize that this basis does not need to be exponentially-localized, as this is not always possible, such as for quantum Hall systems~\cite{rashba97}.

% projectors
We then define a family of projectors $P_\rho : \rho\in[0,1]$, which remove a fraction $1-\rho$ of the maximally-localized single-particle basis, such that
\begin{equation}
\label{eq:sp_proj}
P_\rho = P_\text{band} - \sum_{i\in\mathcal{L}_\rho} \ket{\tilde{\psi}_i}\bra{\tilde{\psi}_i},
\end{equation}
where $P_\text{band}$ is a projector onto the relevant single-particle band, and $\ket{\tilde{\psi}_i}$ is the symmetrically-orthogonalized wavefunction at site $i$ in the removal subregion $\mathcal{L}_\rho$, which satisfies $|\mathcal{L}_\rho|\leq|\mathcal{L}_0|\leq|A|$.~\footnote{Note that the size of the removal subregion corresponding to a complete set of states, $|\mathcal{L}_0|$, is not necessarily equal to the size of the region $|A|$, as shown for Landau levels in Sec.~\ref{subsubsec:ll_model}.} We note that although the states in the basis are generally linearly independent, they are not mutually orthogonal, and so we first need to symmetrically orthogonalize the removal states, such that $\{\ket{\psi}\}\to\{\ket{\tilde{\psi}}\}$. Details of the symmetric orthogonalization procedure are given in Sec.~SI of the Supplementary Material. Crucially, since the quasilocal projector $\ket{\tilde{\psi}_i}\bra{\tilde{\psi}_i}$ corresponds to the number operator $n_i$ for the single-particle orbital at site $i\in\mathcal{L}_\rho$, the overall projector $P_\rho$ restricts the system to a Hilbert space in which these operators have a fixed (zero) eigenvalue. Hence, by eliminating a maximally-localized orthogonal subset of states in the basis, we truncate the Hilbert space $\mathcal{H}\to\mathcal{H}'$.

% renormalization
Provided the family of projectors $P_\rho$ are statistically self-similar under this decimation, we can relate the truncated system $\bar{H}(r)=P_\rho H(\mathbf{r}) P_\rho$ to the original system $H(\mathbf{r})$ by a rescaling $\mathbf{r'}=b\mathbf{r}$ and renormalization $H'(\mathbf{r}')=\zeta^{-1} \bar{H}(\mathbf{r}')$. In this way, we can consider the ground state of $H'(\mathbf{r}')$ as the new ground state defined on the space of states\footnote{Similarly, in the case of multi-component systems, we can choose a composite wavefunction basis $\{\ket{\psi^\text{c}}\}$ to define the procedure.}. We then iterate this process to construct a renormalization flow, based on the removal of maximally-localized single-particle orbitals, which we call ``localization renormalization". 

% summary
For the case of quantum Hall systems, we deduce the rescaling and renormalization factors to be $b=\rho^{-1/2}$ and $\zeta=\rho$, respectively. In particular, we observe a divergence of the localization length in the $\rho\to 0, L\to\infty$ limits, governed by a universal scaling exponent $\nu\sim 1/2$ for topological bands, whereas we observe convergence to a system-size-independent constant in the trivial phase. This holds independently of how the single-particle states are removed and how the localization length is defined.

%%%%%%%%%%%%%%%%%%%%%%%%%%%%%%%%%%%%%%%%%%%%%
\section{Quantum Hall Examples} %
\label{sec:qh_ex}                           %
%%%%%%%%%%%%%%%%%%%%%%%%%%%%%%%%%%%%%%%%%%%%%

In this section, we demonstrate the localization renormalization procedure through the use of two examples, based on the integer quantum Hall effect. In Sec~\ref{subsec:ll}, we examine continuous systems in the form of Landau levels, and in Sec.~\ref{subsec:ci}, we examine discrete systems in the form of Chern insulators.

\subsection{Landau Levels}
\label{subsec:ll}

To begin, we focus on Landau levels. In Sec.~\ref{subsubsec:ll_model}, we summarize the Landau level Hamiltonian and properties of coherent states, in Sec.~\ref{subsubsec:ll_method}, we explain the state removal algorithm, and in Sec.~\ref{subsubsec:ll_results}, we present numerical results for the localization renormalization.

\subsubsection{Model}
\label{subsubsec:ll_model}

We consider a free spinless electron of mass $m_\text{e}$ and charge $-e$, confined to the $xy$-plane, in the presence of a perpendicular magnetic field $\mathbf{B}=B\hat{\mathbf{e}}_z$. The Hamiltonian is given as $H_\text{LL}=\boldsymbol{\pi}^2/2m_\text{e}$, where $\boldsymbol{\pi}=\mathbf{p}-e\mathbf{A}$ is the dynamical momentum, $\mathbf{p}$ is the canonical momentum, and $\mathbf{A}$ is the vector potential. Since $\pi_x$ and $\pi_y$ are canonical conjugates, this Hamiltonian has the same structure as a harmonic oscillator, such that $H_\text{LL}=\hbar\omega_\text{c}(a^\dagger a + 1/2)$, where $\omega_\text{c}$ is the cyclotron frequency and $a^{(\dagger)}$ are the ladder operators hopping between energy levels. The eigenspectrum is composed of evenly-spaced and highly-degenerate Landau levels at energies $E_n = \hbar \omega_\text{c}(n+1/2)$, where $n$ is the Landau level index. Using symmetric gauge, we may simply express the angular momentum operator as $L_z=\hbar(a^\dagger a - b^\dagger b)$, where we have introduced the ladder operators $b^{(\dagger)}$, defined using center coordinates $\mathbf{R}$ conjugate to $\boldsymbol{\pi}$, governing the angular momentum quantum number $m$. Hence, the Landau level states are conventionally indexed as $\ket{n,m}$~\cite{yoshioka98, AndrewsThesis}.

The coordinate representation of a Landau level wavefunction may be obtained by solving the differential equation $a\ket{0,m}=b\ket{n,0}=0$ in an appropriate basis, and can be subsequently translated through a distance $\boldsymbol{\delta}$ using the magnetic translation operator $t(\boldsymbol{\delta})=\exp(-\mathrm{i}\boldsymbol{\delta}\cdot\mathbf{K}/\hbar)$, where $\mathbf{K}=\mathbf{p}-e\mathbf{A}+e\mathbf{B}\times\mathbf{r}$ is the pseudomomentum, which commutes with the Hamiltonian. In symmetric gauge, the Landau level states take the form of a Gaussian, modulated by Laguerre polynomial and monomial prefactors~\cite{yoshioka98}. From the eigenspectrum of the angular momentum operator, we can see that the most-localized Landau level states are obtained at $n=m$ with second moment $\braket{n,n|r^2|n,n}=2(2n+1)\ell^2$, where $\ell$ is the magnetic length. For example, the most-localized state in the lowest Landau level (LLL) is given in a coordinate representation as $\phi_{0,0}(\mathbf{r})\sim\exp(-r^2/4\ell^2)$ with a second moment $\braket{0,0|r^2|0,0}=2\ell^2$.

In this section, we focus our attention on coherent states, which are non-dispersive wavepackets that saturate the uncertainty principle and correspond to classical observable evolution. Formally, these states are defined as eigenstates of the lowering operator. Since we restrict ourselves to a single Landau level, we consider the eigenstates of $b\ket{\beta}=\beta\ket{\beta}$, where $\ket{\beta}$ is a coherent state with corresponding eigenvalue $\beta$. Coherent states have several important properties for this study. In particular, since the ladder operators $b^{(\dagger)}=\frac{1}{\sqrt{2}}(X+\mathrm{i}Y)^{(\dagger)}$ are defined using the center coordinates $\mathbf{R}$, saturating the Heisenberg uncertainty principle corresponds to $\Delta X \Delta Y = \hbar/2$. Hence, a coherent state in a Landau level is necessarily a maximally-localized state and so may be obtained by magnetically translating the $n=m$ Landau level states, described above~\cite{malkin69}. Removing maximally-localized states in Landau levels is the main step of our Hilbert space truncation algorithm, outlined in Sec.~\ref{subsubsec:ll_method}.

Finally, we note that coherent states trivially form an overcomplete basis for a given Landau level, since they are enumerated by an uncountable set, whereas the orthogonal basis of angular momentum eigenstates is countable. Therefore, it is desirable to truncate the set of coherent states to obtain a complete basis. Indeed, it has been shown that we can form a critical basis set by restricting coherent states to the sites of a grid in the $XY$-plane of unit cell area $S=2\pi$, and an overcomplete set with $S<2\pi$~\cite{perelomov71}\textsuperscript{,}\footnote{From now on, we work in natural units.}\textsuperscript{,}\footnote{In the original paper by Perelomov~\cite{perelomov71}, they work in the space of complex eigenvalues $\beta$ to obtain a critical unit cell area $S=\pi$. However, in real space, the wavepackets are centered at coordinates $\mathbf{r}_{\beta} = \sqrt{2}(\text{Re}(\beta), \text{Im}(\beta))$, which yields a critical unit cell area $S=2\pi$.}. By removing just one state from the critical set, we can make the set exactly complete. However, na\"ively attempting to symmetrically orthogonalize such a complete set of coherent states in a Landau level compromises locality. It has been shown that the resulting states have a high overlap with Gaussian states at short distances from the origin, but an oscillating power-law decay at long distances with a diverging second moment~\cite{rashba97}. This highlights the topology-localization dichotomy: it is impossible to construct exponentially-localized Wannier functions in a topological system.    

\subsubsection{Method}
\label{subsubsec:ll_method}

Having shown that a critical set of maximally-localized Landau level states may be obtained using a grid of coherent states in the $XY$-plane with a unit cell area of $2\pi$, we can utilize this in our Hilbert space truncation algorithm. In this procedure, we simultaneously remove the most-localized states on the sites of a square grid, converging from an arbitrarily large unit cell area to the unit cell area corresponding to this critical set. However, we note that the results hold independently of the truncation algorithm and removal grid geometry, as demonstrated in Secs.~SII and SIII of the Supplementary Material. The method is as follows.

\begin{figure}
	\centering
	\includegraphics[width=\linewidth]{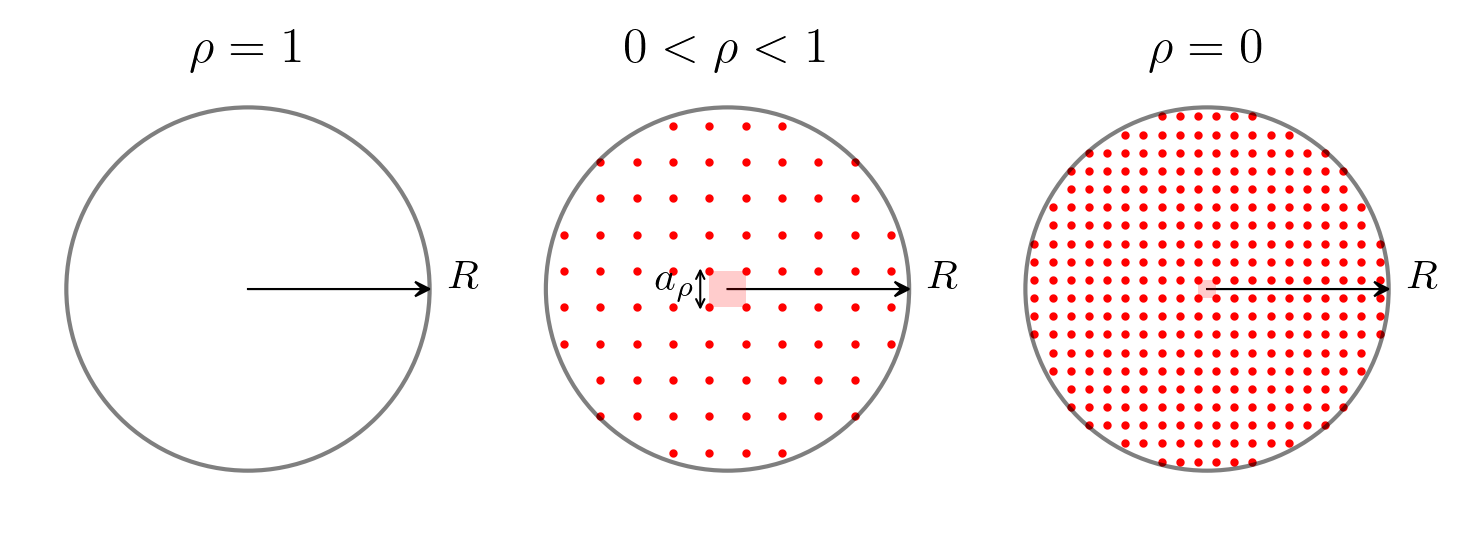}
	\caption{\label{fig:disc_grids} \textbf{Removal lattice for Landau levels.} Convergence of the removal lattice $\mathcal{L}_\rho$ in Landau levels, sketched for three values of $\rho$. The area of the $\mathcal{L}_\rho$ unit cell $a_\rho^2 = 2\pi/(1-\rho)$ is shaded pink, and the disc radius $R$ is marked with an arrow.}
\end{figure}

We work in a Landau level $\ket{n,m}$ defined on a continuous disc of radius $R$, centered at the origin. In each Landau level $n$, we use a truncated angular momentum basis $m\in\{ 0, 1, \dots, \frac{3}{4}R^2 \}$. The basis cut-off is chosen based on empirical convergence, such that the critical grid of wavepackets on the disc is approximately homogeneous, and not distorted by the boundary. Next, we introduce a removal lattice $\mathcal{L}_{\rho}$ with sites on a square grid
\begin{equation}
\label{eq:lattice}
\mathbf{r}_{ij} = a_{\rho}((i-1/2)\hat{\mathbf{e}}_x + (j-1/2)\hat{\mathbf{e}}_y),
\end{equation}
where $a_{\rho}=\sqrt{2\pi / (1-\rho)}$ is the lattice constant and $i,j\in\mathbb{Z}$, as sketched in Fig.~\ref{fig:disc_grids}. The quantity $\rho$ is the fraction of states remaining relative to $\mathcal{L}_0$, defined as $\rho=1-A_0 / A_{\rho}\in[0, 1]$. For each lattice site $\mathbf{r}_{ij}\in\mathcal{L}_{\rho}$, we then find the maximally-localized state $\ket{\psi_{ij}}$, which in this case is given by the Landau level coherent state.

This defines the set of maximally-localized states in a Landau level for each $\rho$. However, although these states are generally linearly independent, they are not mutually orthogonal. Therefore, in order to project these states out of the Hilbert space, we first need to orthogonalize the subspace. To this end, we use the symmetric orthogonalization procedure to transform $\{\ket{\psi}\}\to\{\ket{\tilde{\psi}}\}$. We can then apply the projector 
\begin{equation}
\label{eq:proj}
P^\text{LL}_{\rho} = P_{n\text{LL}} - \sum_{i,j\in\mathcal{L}_{\rho}} \ket{\tilde{\psi}_{ij}}\bra{\tilde{\psi}_{ij}},
\end{equation}
where $P_{n\text{LL}}$ is the projector to the $n$th Landau level.

Finally, on each removal iteration we record the localization length of the system $\xi$. Note that we have offset the removal lattice from the origin, as shown in Eq.~\eqref{eq:lattice}, so that we can use this as our reference site for the most-localized state in the system. The origin is also furthest away from the boundary of the disc and hence least susceptible to finite-size effects. Although there are many metrics for quantifying the localization length, we choose the minimum eigenvalue of the distance-squared matrix~\cite{marzari97}, such that $D^2 = P^\text{LL}_{\rho} \mathbf{r}^2 P^\text{LL}_{\rho}$. A discussion of metrics for the localization length is presented in Sec.~SIV of the Supplementary Material.

In summary, we identify the maximally-localized states $\{\ket{\psi}\}$ at sites $\mathbf{r}_{ij}\in\mathcal{L}_{\rho}$, after which we symmetrically orthogonalize the states $\{\ket{\psi}\}\to\{\ket{\tilde{\psi}}\}$, project them out of the system using $P^\text{LL}_\rho$, and subsequently compute the localization length $\xi$, corresponding to the second moment of the maximally-localized state at the origin. We start at $\rho\lesssim 1$ and then repeat for decreasing $\rho$, taking the limit $\rho \to 0 $, which corresponds to eliminating a critical basis of states and hence a localization length divergence.

\subsubsection{Results}
\label{subsubsec:ll_results}

\begin{figure}
	\centering
	\includegraphics[width=\linewidth]{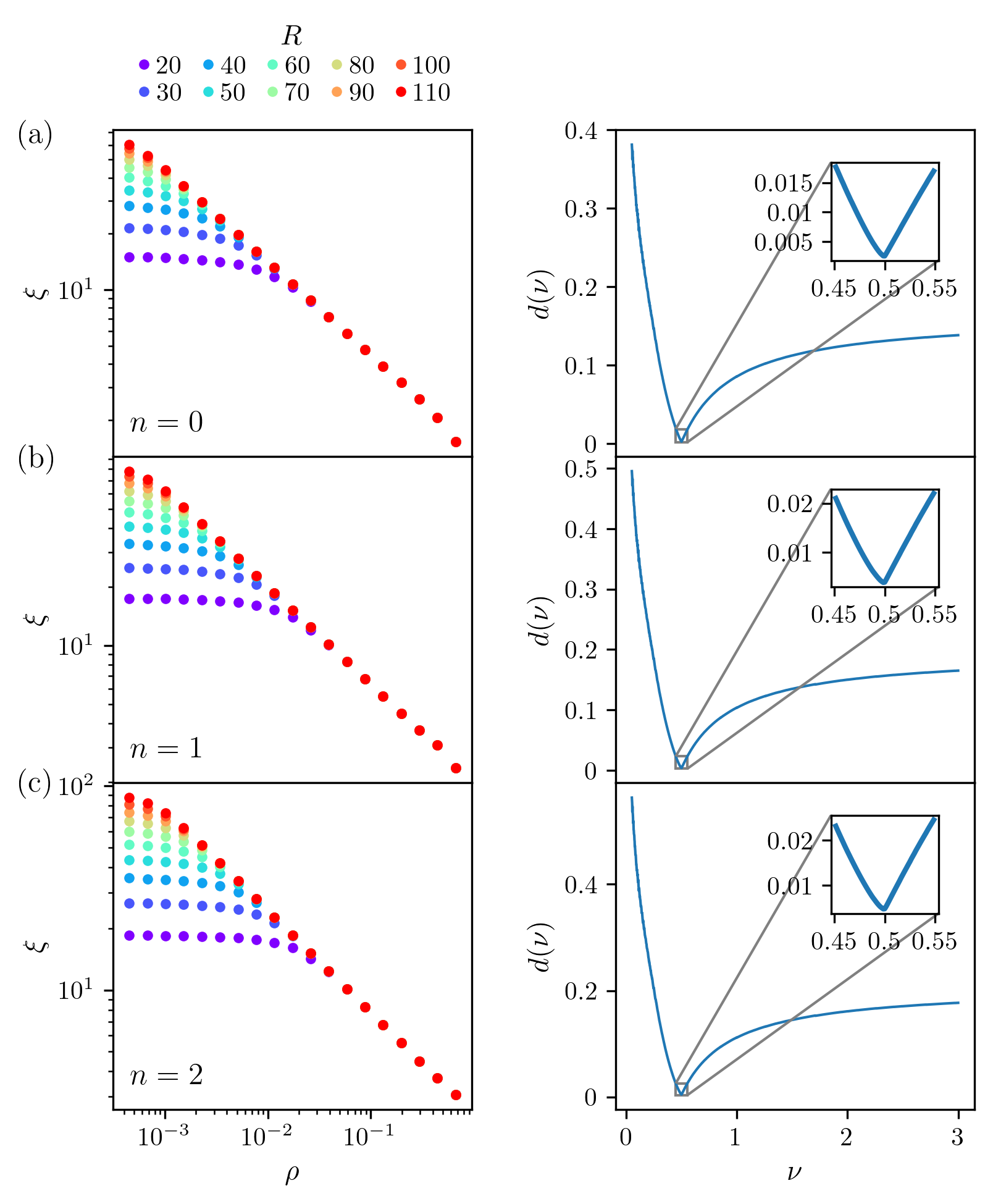}
	\caption{\label{fig:method1_nLL} \textbf{Localization scaling in Landau levels.} (left)~Localization length $\xi(\rho)$ obtained by the simultaneous elimination of a lattice of localized states in the (a)~$n=0$, (b)~$n=1$, and (c)~$n=2$ Landau levels. The system size $R$ is depicted in different colors. At small $\rho$ and large system sizes, the relationship is linear indicating power-law behavior in the thermodynamic limit. (right)~The quality function of data collapse $d(\nu)$ for the (a)~$n=0$, (b)~$n=1$, and (c)~$n=2$ Landau levels; optimization of the quality function results in critical exponents of (a)~$\nu_{\text{0LL}}=0.500\pm0.005$, (b)~$\nu_{\text{1LL}}=0.500\pm0.006$, and (c)~$\nu_{\text{2LL}}=0.500\pm0.006$.}
\end{figure}

After iteratively performing this state removal, we can plot the behavior of $\xi$ as a function of $\rho$ in each case. In the left panels of Fig.~\ref{fig:method1_nLL}, we show the scaling of $\xi(\rho)$ at various system sizes $R$, in the (a)~$n=0$, (b)~$n=1$, and (c)~$n=2$ Landau levels. From the figures, we can see a divergence of the localization length in the $\rho\to 0 $ and $R\to\infty$ limits.\footnote{Note that the order of limits in not significant.} In order to quantify this divergence, we perform a finite-size scaling analysis. By finding a dimensionless scaling function $g$, which satisfies
\begin{equation}
\label{eq:thermo}
\xi(\rho)/\xi_{\infty}(\rho)=g(R/\xi_{\infty}(\rho)),
\end{equation}
where $\xi_{\infty}$ is the extrapolated localization length in the thermodynamic limit, we can verify the power-law divergence $\lim_{\rho\to 0}\xi(\rho)\sim\rho^{-\nu}$. Furthermore, by quantifying how well the data from Eq.~\eqref{eq:thermo} collapse onto the same curve, we can compute the scaling exponent $\nu$~\cite{bhattacharjee01}. Minimizing a quality metric of data collapse $d(\nu)$~\cite{nelder65}, yields the scaling exponent in each case, as shown in the right panels of Fig.~\ref{fig:method1_nLL}. The computed values, $\nu_{0\text{LL}}=0.500\pm0.005$, $\nu_{1\text{LL}}=0.500\pm0.006$, and $\nu_{2\text{LL}}=0.500\pm0.006$, accord precisely with a scaling exponent of $\nu=0.5$ in each case, with only minor discrepancies due to numerical imprecision. Here, the error bars are given as the $1\%$ fluctuations of the quality metric about the minimum, computed using the Hessian $d''(\nu)$. Further details on the finite-size scaling analysis are given in Sec.~SV of the Supplementary Material.

\begin{figure}
	\centering
	\includegraphics[width=\linewidth]{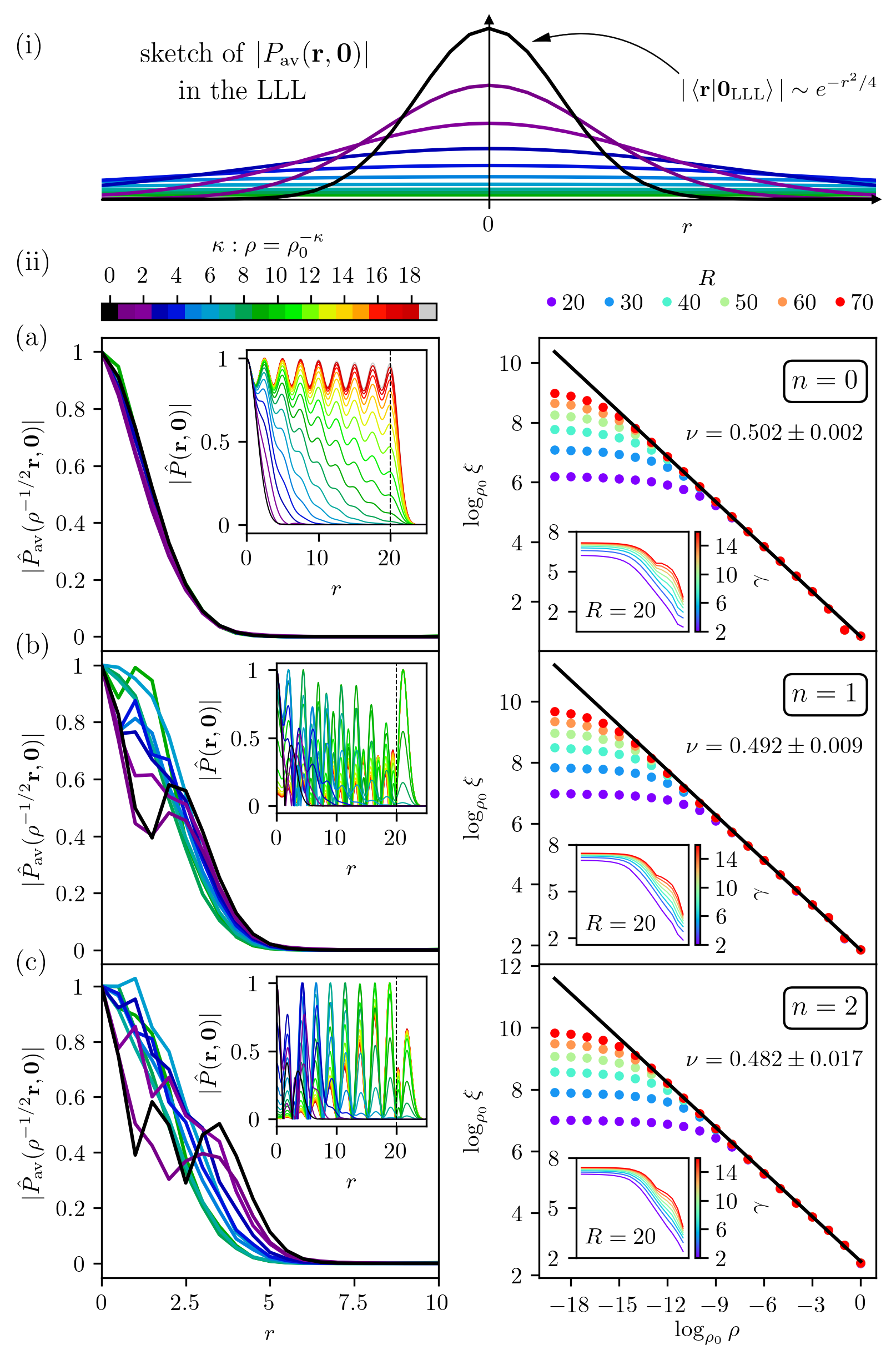}
	\caption{\label{fig:method1_nLL_proj} \textbf{Projector expansion in Landau levels.} (i)~Sketch of the averaged real-space projector expansion as $\rho\to 0$ in the LLL. (ii)~(left)~Real-space radial profile of the projector $P(\mathbf{r}, \mathbf{0})=\braket{\mathbf{r}|P^\text{LL}_{\rho}|\mathbf{0}}$. The rotationally- and translationally-averaged projector, normalized with respect to the initial value, $\hat{P}_\text{av}$, is shown in the main plot. The original unaveraged projector, normalized with respect to the maximum value, $\hat{P}$, is shown inset. The projectors are presented for the (a)~$n=0$, (b)~$n=1$, and (c)~$n=2$ Landau levels, at a system size of $R=40$ in the main plot and $R=20$ in the inset, with $\rho_0=1.5$. The boundary of the system at $r=20$ in the inset is marked with a dashed line. Note that the projectors are plotted in reverse order, such that the $\kappa=0$ line is on top. (ii)~(right)~Finite-size scaling of the second moment of the projector for the (a)~$n=0$, (b)~$n=1$, and (c)~$n=2$ Landau levels. The lines of best fit, for the linear region of the largest system size, are overlaid in black. The corresponding higher moments, $\xi^\gamma=\braket{r^\gamma}$, are shown inset for $R=20$.}	
\end{figure}

To gain further insight on the nature of this scaling, we examine the projector $P^\text{LL}_\rho$ at each step, as sketched in Fig.~\ref{fig:method1_nLL_proj}(i). In the left panels of Fig.~\ref{fig:method1_nLL_proj}(ii), we show the magnitude of the projector in real space, relative to the origin, $|P(\mathbf{r}, \mathbf{0})| = |\braket{\mathbf{r}|P^\text{LL}_{\rho}|\mathbf{0}}|$ in the (a)~$n=0$, (b)~$n=1$, and (c)~$n=2$ Landau levels. Starting with the inset of Fig.~\ref{fig:method1_nLL_proj}(ii)(a), we present the normalized projector $|\hat{P}|\sim\rho^{-1}|P|$ for various $\rho$ in the smallest system size $R=20$. At $\rho=1$, we start with the Gaussian state at the origin, and on each removal iteration the projector expands, continuing until the edge of the system ($r=20$) is reached at $\rho\approx1.5^{-9}$. This corresponds to the value of $\rho$ where the linear scaling breaks down in Fig.~\ref{fig:method1_nLL}(a). We note also that there are slight oscillations in the amplitude of $\hat{P}$ with length scale $a_0=\sqrt{2\pi}$, corresponding to the modulation of the removal lattice $\mathcal{L}_0$. Crucially, in the right panel of Fig.~\ref{fig:method1_nLL_proj}(ii)(a), we plot the localization length quantified via the second moment of the projector against $\rho$, and show that we can recover the same scaling exponent. In the inset of the right panel of Fig.~\ref{fig:method1_nLL_proj}(ii)(a), we further show that this scaling relation is not only recovered from the second moment but also for higher moments, $\gamma \lesssim 10$\footnote{We note that a local maximum emerges as we increase $\gamma\gtrsim 10$, which breaks the linear scaling at large $\rho$. Although it is difficult to interpret higher-order moments visually, this effect may be due to the critical lattice modulation causing a proportionally larger distortion of projectors with smaller radii.}. This indicates that the projector is self-similar in a statistical sense\footnote{Self-similarity is mathematically defined for all moments, so that distributions are statistically identical. However, in practice, the definition is often taken to refer to the first two moments only~\cite{Embrechts00, Peng00, Mandelbrot10, Andrews22}.}. Further details of the moment computations are provided in Sec.~SVI of the Supplementary Material. Finally, to elucidate this statistical self-similarity, we can rescale the family of projectors $P^\text{LL}_\rho$ to show that they collapse onto the same curve. In the left panel of Fig.~\ref{fig:method1_nLL_proj}(ii)(a), we present the translationally- and rotationally-averaged normalized projector $\hat{P}_\text{av}$, which eliminates any artifacts specific to the removal lattice, and we plot this using the rescaled spatial coordinates $\mathbf{r}'=\rho^{-1/2}\mathbf{r}$. Here, we clearly observe that the projectors collapse onto the initial Gaussian. In higher Landau levels, this picture also holds, albeit slightly obscured by the complexity of our initial coherent state $|\braket{\mathbf{r}|\mathbf{0}_{n\text{LL}}}|\sim e^{-r^2/4}|L_n (r^2 / 2)|$, where $L_n$ is the $n$th Laguerre polynomial. Again, we find that the second moment of the expanding projector can be used as a proxy for the localization to extract the scaling exponent, as shown in the right panels of Figs.~\ref{fig:method1_nLL_proj}(ii)(b,c). Moreover, the projector is statistically self-similar up to high moments $\gamma\lesssim 10$, as shown in the insets. Rescaling the projector, as before, we find that the curves approximately collapse onto a single Gaussian, as shown in the left panels of Figs.~\ref{fig:method1_nLL_proj}(ii)(b,c), however this is now more difficult to discern visually due to the Laguerre polynomial modulation in the initial coherent state. In all cases, we obtain a scaling exponent $\nu\simeq0.5$ from the self-similar projector expansion. Although this is numerically more challenging to extract for higher Landau levels, all three computed values agree within error bars.                  

\subsection{Chern Insulators}
\label{subsec:ci}

To complement this analysis, we now focus on Chern insulators. In Sec.~\ref{subsubsec:ci_model}, we define the tight-binding model, in Sec.~\ref{subsubsec:ci_method}, we outline the truncation algorithm for discrete systems, and in Sec.~\ref{subsubsec:ci_results}, we present results for the scaling exponent.

\subsubsection{Model}
\label{subsubsec:ci_model}

We consider a free spinless electron confined to a lattice on the $xy$-plane. One of the most well-understood examples of a Chern insulator is the Haldane model~\cite{haldane88}, defined on the honeycomb lattice as
\begin{equation}
\begin{split}
H_{\text{CI}} = -t_1 \sum_{\braket{ij}} c_i^\dagger c_j - t_2 \sum_{\braket{\braket{ij}}} e^{\pm\mathrm{i}\phi} c_i^\dagger c_j \\
+ M \sum_i (n_{A,i}-n_{B,i}) + \text{H.c.},
\end{split}
\end{equation}
where $t_1$ and $t_2$ are the amplitudes corresponding to nearest- ($\braket{ij}$) and next-nearest-neighbor ($\braket{\braket{ij}}$) hoppings, $c^{(\dagger)}$ are the spinless fermion creation(annihilation) operators, $e^{\pm\mathrm{i}\phi}$ is the next-nearest-neighbor complex phase factor, $M$ is the staggered chemical potential, and $n_{A(B)}$ is the density operator on sublattice A(B). The sign of the complex phase is determined by the direction of the next-nearest-neighbor hopping. The phase is positive clockwise around a minimal down-pointing triangle of the A sublattice and anti-clockwise around a minimal down-pointing triangle of the B sublattice. By carefully selecting the parameters in the Haldane model, we can tune between a topological and trivial phase.

\begin{figure}
	\centering
	\includegraphics[width=\linewidth]{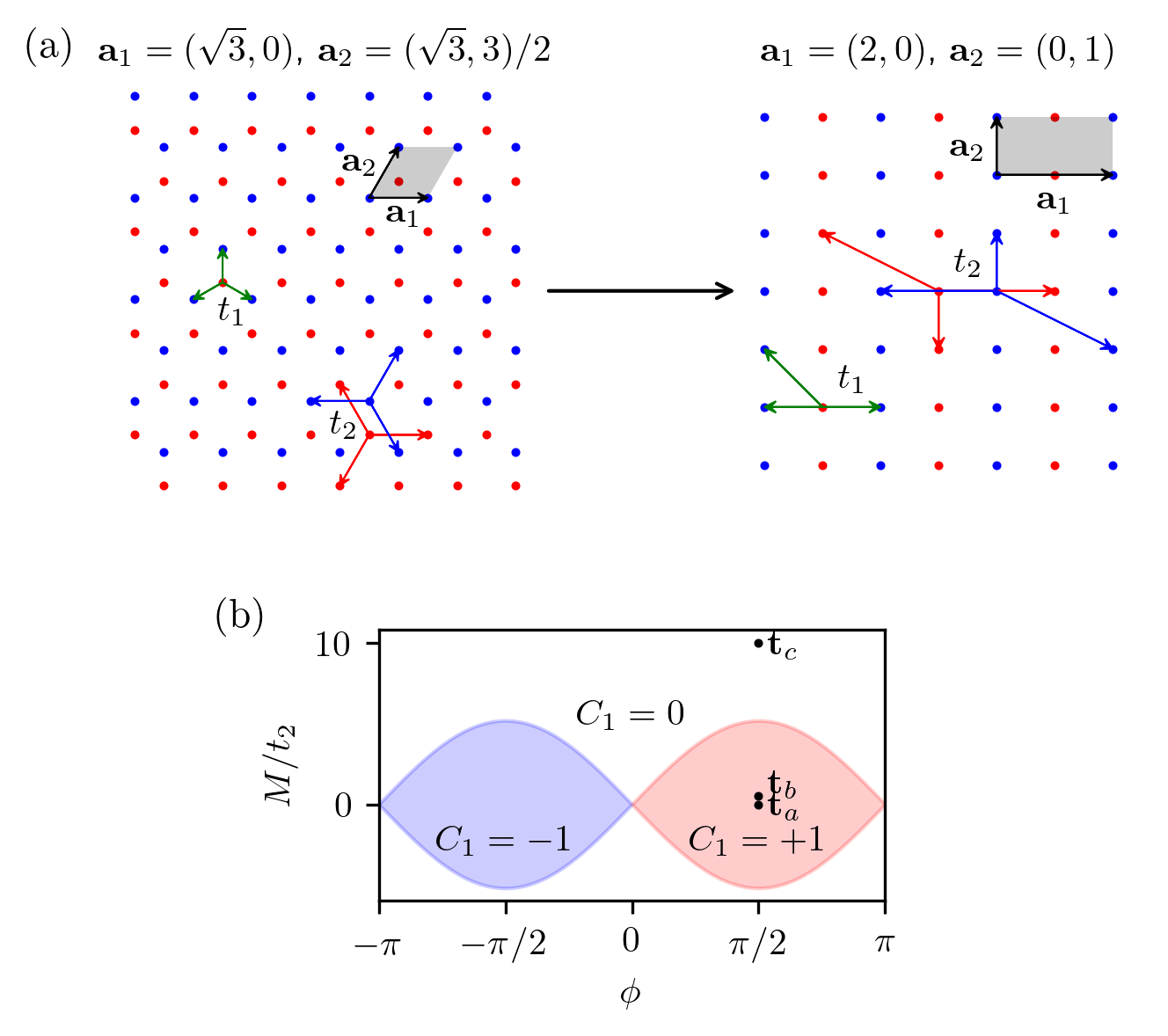}
	\caption{\label{fig:square_Haldane} \textbf{Haldane model on a square lattice.} (a)~Mapping of the Haldane model onto a square lattice. The basis vectors $\{\mathbf{a}_1, \mathbf{a}_2\}$ are given in units of the lattice constant and the unit cell is shaded gray. The $t_1$ hoppings are colored green, and the $t_2$ hoppings are colored red(blue) according to their A(B) sublattice, with the direction of the arrows corresponding to a positive complex phase. (b)~Phase diagram for the Haldane model, as a function of complex phase $\phi$ and chemical potential $M$, colored according to the sign of the Chern number for the lowest band $C_1$. The three selected parameter sets are at $\mathbf{t}_a=\{t_2=0.1, M=0\}$, $\mathbf{t}_b = \{t_2=0.2, M=0.1\}$, and $\mathbf{t}_c = \{t_2=0.1, M=1\}$. For all parameter sets, $\phi=\pi/2$, and in all cases, we set $t_1=1$.}
\end{figure}

For consistency with our state removal algorithm defined for Landau levels in Sec.~\ref{subsubsec:ll_method}, we map the Haldane model onto a square lattice with orthogonal basis vectors $\mathbf{a}_1 \cdot \mathbf{a}_2 = 0$, as depicted in Fig.~\ref{fig:square_Haldane}(a). Although the geometry of the model changes under this mapping, with nearest- and next-nearest neighbors no longer preserved, the topology is unchanged~\cite{Simon20}. Specifically, we obtain the same Haldane lobe phase diagram in the $M/t_2$ against $\phi$ parameter plane. The Haldane Hamiltonian yields a two-band eigenspectrum, where the bottom band has Chern number $C_1=-1$ for $|M/t_2|\leq 3\sqrt{3}\sin(\phi)$ with $\phi\in(-\pi,0)$, $C_1=1$ for $|M/t_2|\leq 3\sqrt{3}\sin(\phi)$ with $\phi\in(0,\pi)$, and $C_1=0$ elsewhere. In this section, we restrict ourselves to the physics of the lowest band and study three configurations, $\mathbf{t}_a$ (topological), $\mathbf{t}_b$ (topological), and $\mathbf{t}_c$ (trivial), as shown in Fig.~\ref{fig:square_Haldane}(b). Further details on the choice of parameters are discussed in Sec.~SVII of the Supplementary Material. We note that the localization renormalization described in this section works for any choice of Chern insulator. We choose the Haldane model due to its popularity; motivated particularly by recent realizations in moir{\'e} materials~\cite{Zhao22} and theoretical studies of its topological phase transitions~\cite{Mildner23}.  

\subsubsection{Method}
\label{subsubsec:ci_method}

The method is similar to the procedure described for Landau levels in Sec.~\ref{subsubsec:ll_method}. However, there are a few technical nuances specific to discrete systems, that increase the complexity in this case. As before, the renormalization is independent of the details of the algorithm, as shown in Secs.~SII and SIII of the Supplementary Material. The method is as follows.

\begin{figure}
	\centering
	\includegraphics[width=\linewidth]{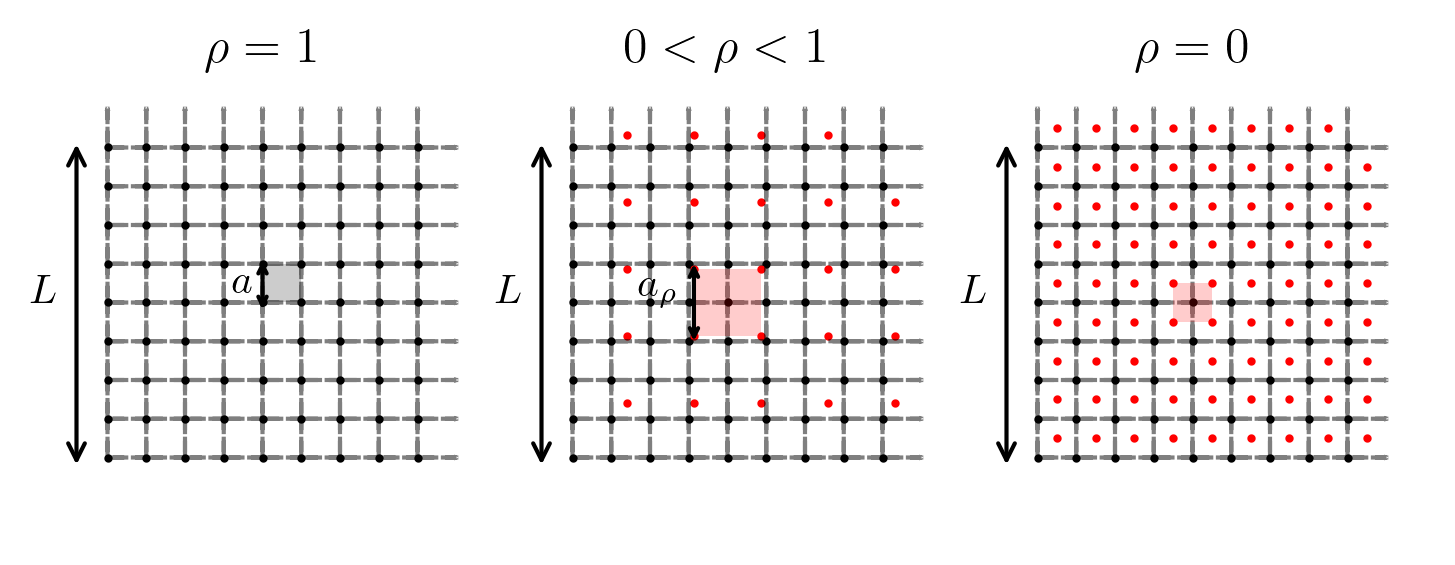}
	\caption{\label{fig:torus_grids} \textbf{Removal lattice for the Haldane model.} Convergence of the removal lattice $\mathcal{L}_\rho$ for the square-lattice Haldane model, sketched for three values of $\rho$. The area of the $\mathcal{L}$ unit cell $a^2=1$ is shaded gray, the area of the $\mathcal{L}_\rho$ unit cell $a_\rho^2 = 1/(1-\rho)$ is shaded pink, and the linear system size $L$ is marked with an arrow.}
\end{figure}

We start with an underlying square lattice for the Haldane model $\mathcal{L}_\text{H}$, with lattice constant $a=1$, dimensions $L\times L$, and periodic boundary conditions. Here, the simple translational symmetry and toroidal topology are expedient for numerical simulations. Next, we introduce a square removal lattice $\mathcal{L}_\rho$, with lattice constant $a_\rho=\sqrt{1/(1-\rho)}$, symmetrically offset from a chosen ``origin" site of $\mathcal{L}_\text{H}$, analogously to how $\mathcal{L}_\rho$ was offset on the disc in Eq.~\eqref{eq:lattice}, as sketched in Fig.~\ref{fig:torus_grids}. For each site of the removal lattice $\mathbf{r}_{ij}\in\mathcal{L}_\rho$, we then find the corresponding most-localized state $\ket{\psi_{ij}}$ in the $\mathcal{L}_\text{H}$ basis. In contrast to the continuous case, since $\mathcal{L}_\text{H}$ and $\mathcal{L}_\rho$ are incommensurate lattices, we can no longer exploit conventional translation operators to simplify the computations. Instead, we compute the most-localized state at each $\mathbf{r}_{ij}\in\mathcal{L}_\rho$, by extracting the minimum-eigenvalue state of the distance-squared matrix $D^2_{ij} = P_{\text{LB}} (\mathbf{r}-\mathbf{r}_{ij})^2 P_{\text{LB}}$, where $P_{\text{LB}}$ is the projector to the lowest band. Furthermore, care needs to be taken when $\mathbf{r}_{ij}$ falls on an axis of $\mathcal{L}_\text{H}$, since this can result in a 2- or 4-fold degeneracy for the minimum localization length. In these cases, we break the degeneracy to avoid any linear dependencies by selecting the state that has the smallest overlap with the previously selected states $\{\ket{\psi}\}$.  Finally, due to the incommensurability of $\mathcal{L}_\text{H}$ and $\mathcal{L}_\rho$, coupled with the periodic boundary conditions, we need to ensure that $\mathcal{L}_\rho$ does not have any overlapping regions.

Now that we have defined the set of maximally-localized states in our Chern insulator for each removal iteration $\rho$, we can proceed to remove these states from the system. As before, we note that although the most-localized states corresponding to our removal lattice $\mathcal{L}_\rho$ are generally linearly independent, they are not mutually orthogonal. Hence, we first symmetrically orthogonalize the set of maximally-localized states $\{\ket{\psi}\}\to\{\ket{\tilde{\psi}}\}$ and then project them out of the system, using
\begin{equation}
P^\text{CI}_{\rho} = P_{\text{LB}} - \sum_{i,j\in\mathcal{L}_{\rho}} \ket{\tilde{\psi}_{ij}}\bra{\tilde{\psi}_{ij}}.
\end{equation}
Note that this has the same form as the projector in Eq.~\eqref{eq:proj}\footnote{Although the dimension of $P_{\rho}^\text{CI}$ is $2L^2$, its rank is at most $L^2$ due to the band projection.}. Finally, we record the localization length, which we quantify as the minimum eigenvalue of the distance-squared operator, with respect to the origin $D^2=P^\text{CI}_\rho \mathbf{r}^2 P^\text{CI}_\rho$. In this case, care needs to be taken to ensure that we are minimizing distances on the torus. As before, we have designed our system such that the reference ``origin" site hosts the most-localized remaining state and is least-susceptible to numerical artifacts. We start with a large lattice constant with $\rho\lesssim 1$, and then symmetrically shrink the lattice by taking the limit $\rho\to 0$. In this limit, the $\mathcal{L}_\text{H}$ and $\mathcal{L}_\rho$ interpenetrating lattices have the same lattice constant, which corresponds to the removal of a complete basis of states and hence a divergence of the localization length for a topological phase, in general~\cite{Guna23}.     

\subsubsection{Results}
\label{subsubsec:ci_results}

\begin{figure}
	\centering
	\includegraphics[width=\linewidth]{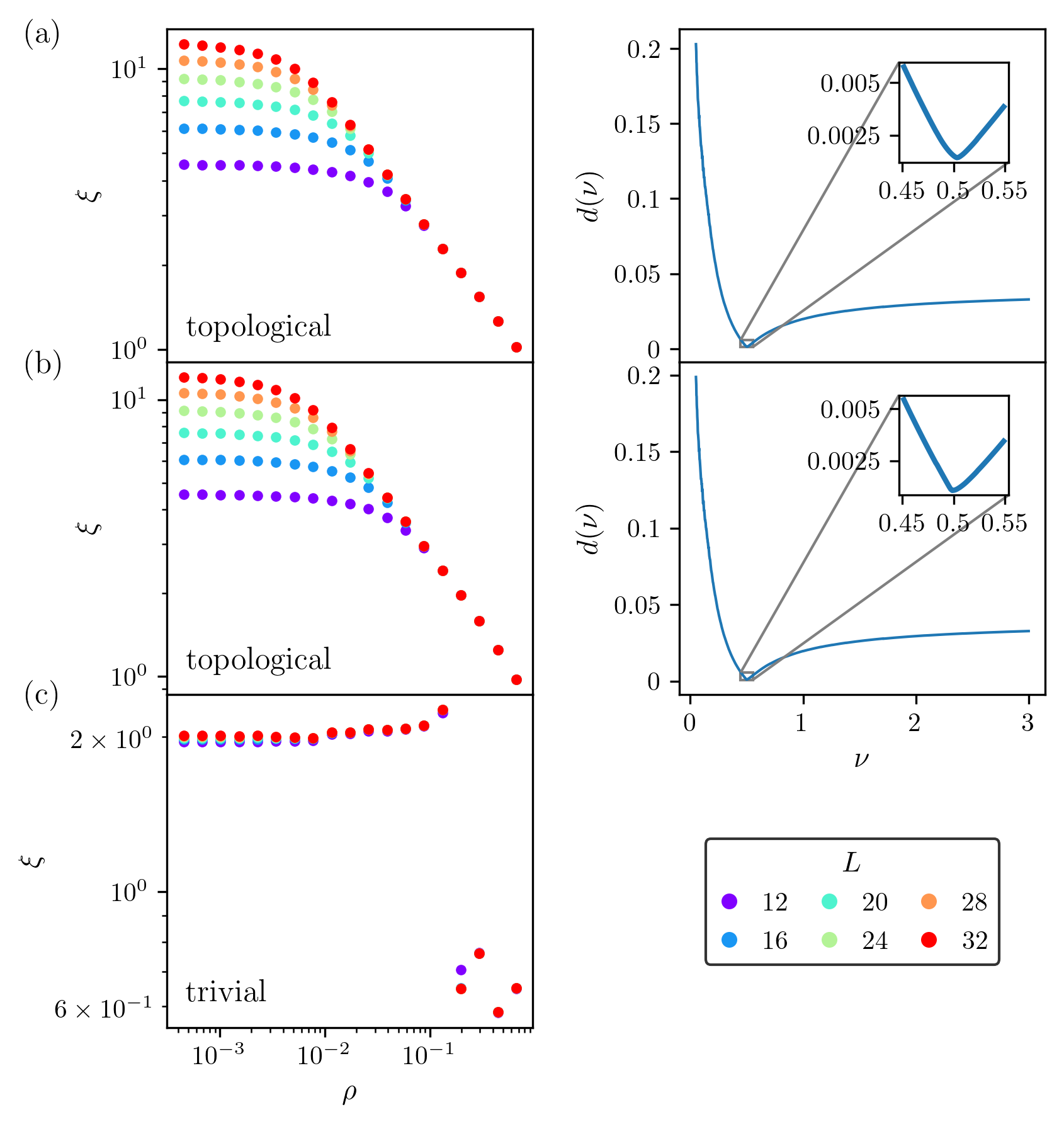}
	\caption{\label{fig:method1_Haldane} \textbf{Localization scaling in the Haldane model.} (left) Localization length $\xi(\rho)$ obtained by the simultaneous elimination of a lattice of localized degrees of freedom, applied to the Haldane model with parameters (a)~$\mathbf{t}_a$, (b)~$\mathbf{t}_b$, and (c)~$\mathbf{t}_c$. The system size $L$ is depicted in different colors. (right) The quality function of data collapse $d(\nu)$ for systems in the topological Haldane model with parameters (a)~$\mathbf{t}_a$ and (b)~$\mathbf{t}_b$; optimization of the quality function results in critical exponents of (a)~$\nu_{\mathbf{t}_a} = 0.503 \pm 0.011$ and (b)~$\nu_{\mathbf{t}_b} = 0.499 \pm 0.012$.}	
\end{figure}

As before, we can plot the localization length $\xi$ as a function of $\rho$ under this iterative state removal. In Fig.~\ref{fig:method1_Haldane}, we present the localization scaling for our three parameter sets: $\mathbf{t}_a$, $\mathbf{t}_b$, and $\mathbf{t}_c$. Starting with the topological configurations shown in Figs.~\ref{fig:method1_Haldane}(a,b), we find a scaling relation similar to that observed for Landau levels in Fig.~\ref{fig:method1_nLL}. That is, the localization length $\xi$ exhibits a power-law divergence as the fraction of states remaining tends to zero $\rho\to 0$ in the thermodynamic limit $L\to \infty$, as shown in the left panels of Figs.~\ref{fig:method1_Haldane}(a,b). Furthermore, using a finite-size scaling ansatz, we can accurately compute the scaling exponent by minimizing a quality metric of data collapse, as shown in the right panels of Figs.~\ref{fig:method1_Haldane}(a,b). Here, we obtain scaling exponents of $\nu_{\mathbf{t}_a}=0.503 \pm 0.011$ and $\nu_{\mathbf{t}_b}=0.499 \pm 0.012$, where the error bars are given by 1\% deviations with respect to the quality metric $d(\nu)$. As before, the scaling exponents take a value of $\nu\simeq0.5$ in each case. Although the precision is reduced due to the discrete nature of the systems, all computed values of $\nu$ from continuous and discrete topological systems agree within errors. In contrast, for the trivial configuration $\mathbf{t}_c$, shown in Fig.~\ref{fig:method1_Haldane}(c), we find that as we decrease $\rho$, the localization length $\xi$ quickly and abruptly converges to an $L$-independent constant\footnote{The localization length converges to a system-size-independent constant following a non-universal readjustment regime, which is sensitive to numerical parameters.}. This shows a sharp distinction to the behavior in topological phases, which supports the use of localization renormalization as an efficient numerical diagnostic for quantum Hall systems.

\begin{figure}
	\centering
	\includegraphics[width=\linewidth]{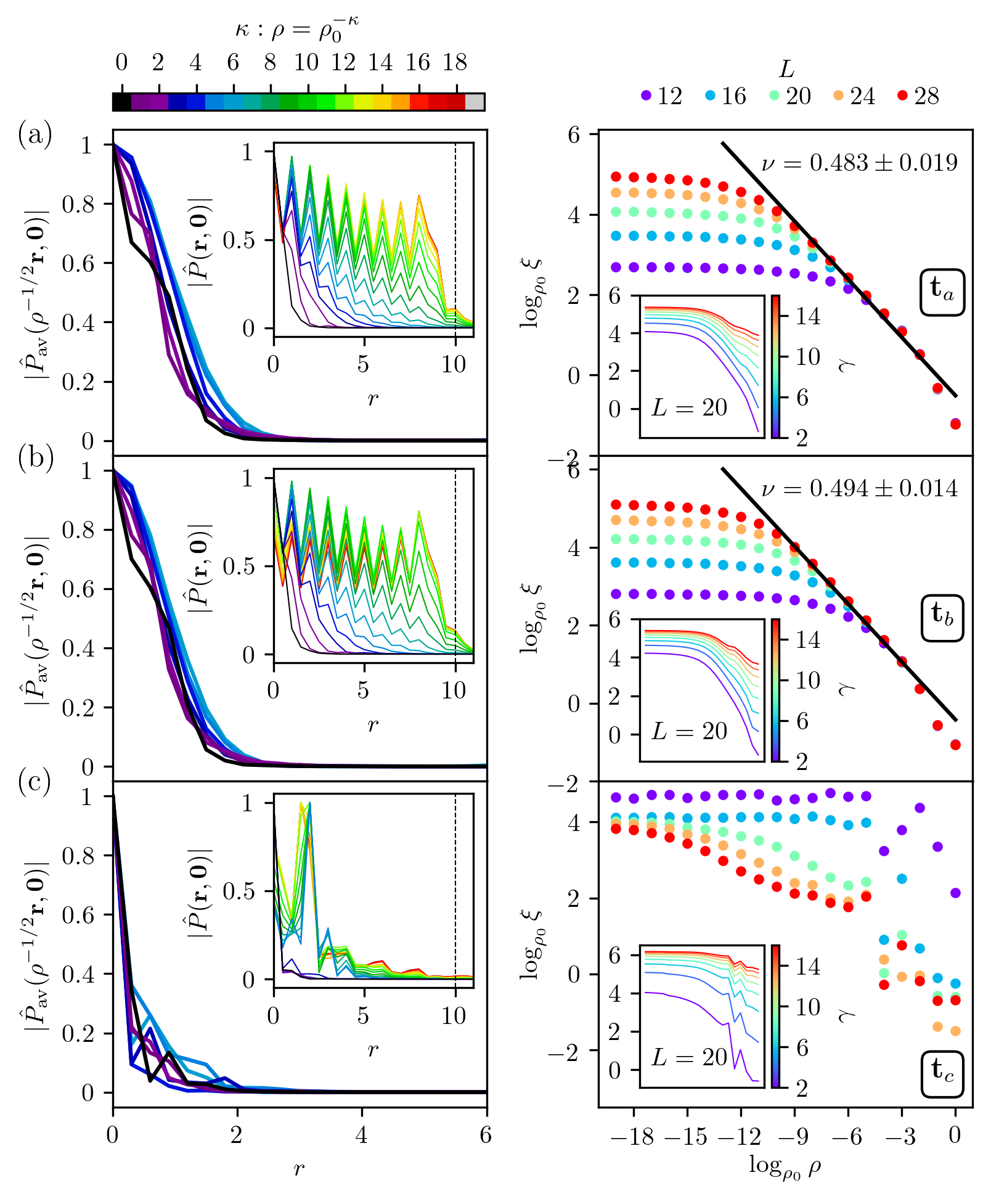}
	\caption{\label{fig:method1_Haldane_proj} \textbf{Projector expansion in the Haldane model.} (left)~Real-space radial profile of the projector $P(\mathbf{r}, \mathbf{0})=\braket{\mathbf{r}|P^\text{CI}_{\rho}|\mathbf{0}}$. The rotationally- and translationally-averaged projector, normalized with respect to the initial value, $\hat{P}_\text{av}$, is shown in the main plot. The original unaveraged projector, normalized with respect to the maximum value, $\hat{P}$, is shown inset. The projectors are presented for the (a)~$\mathbf{t}_a$, (b)~$\mathbf{t}_b$, and (c)~$\mathbf{t}_c$ phases of the Haldane model, at a system size of $L=28$ in the main plot and $L=20$ in the inset, with $\rho_0=1.5$. The boundary of the system at $r=10$ in the inset is marked with a dashed line. Note that the projectors are plotted in reverse order, such that the $\kappa=0$ line is on top. (right)~Finite-size scaling of the second moment of the projector for the (a)~$\mathbf{t}_a$, (b)~$\mathbf{t}_b$, and (c)~$\mathbf{t}_c$ phases of the Haldane model. The lines of best fit, for the linear region of the largest system size, are overlaid in black. The corresponding higher moments, $\xi^\gamma=\braket{r^\gamma}$, are shown inset for $L=20$.}	
\end{figure}

To gain further insight into the distinct scaling behavior in topological and trivial systems, we examine the expansion of the projector in real space. In the left panels of Fig.~\ref{fig:method1_Haldane_proj}, we show the magnitude of the projector $|P(\mathbf{r},\mathbf{0})|=|\braket{\mathbf{r}|P^\text{CI}_\rho|\mathbf{0}}|$ at each removal iteration $\rho$. Starting with the insets in the left panels of Figs.~\ref{fig:method1_Haldane_proj}(a,b), we can see that in the topological phases of the Haldane model, the normalized projector $|\hat{P}|\sim\rho^{-1}|P|$ behaves analogously to the projector in the LLL, shown in the left inset of Fig.~\ref{fig:method1_nLL_proj}(ii)(a). We note that, in this case, the modulation of the normalized projector $\hat{P}$ due to $\mathcal{L}_0$, with length scale $a_0=1$, is exacerbated by the presence of the underlying lattice $\mathcal{L}_\text{H}$. Starting with a Gaussian initial state at $\rho=1$, we observe an expansion of the projector at each removal iteration, until the edge of the system ($r=10$) is reached at $\rho\approx1.5^{-9}$. This corresponds to where the power-law scaling of the $L=20$ curves breaks down in the left panels of Figs.~\ref{fig:method1_Haldane}(a,b). As before, in the right panels of Figs.~\ref{fig:method1_Haldane_proj}(a,b), we show that the second moment of the projector can be used as a proxy for the localization. From this data, we can extract the scaling exponents, which agree with those found in Fig.~\ref{fig:method1_Haldane}, albeit with larger errors due to the indirect nature of the computation. As for Landau levels, we find that this scaling holds not only for the second moment of the projector but also for higher moments, $\gamma\lesssim 10$, indicating statistical self-similarity, as shown in the right insets of Figs.~\ref{fig:method1_Haldane_proj}(a,b). To elucidate this self-similarity, we plot the translationally- and rotationally-averaged normalized projector $\hat{P}_\text{av}$, which eliminates any lattice artifacts, using rescaled coordinates $\mathbf{r}'=\rho^{-1/2}\mathbf{r}$ in the left panels of Figs.~\ref{fig:method1_Haldane_proj}(a,b). Here, we observe that the family of projectors approximately collapse onto the initial Gaussian state. Although this collapse is not as clear as for the LLL case in the left panel of Fig.~\ref{fig:method1_nLL_proj}(ii)(a), due to the spatial discretization, the statistical agreement, in terms of moments, is comparable. In contrast, for the trivial phase of the Haldane model, shown in Fig.~\ref{fig:method1_Haldane_proj}(c), we do not observe a self-similar expansion of the projector in the left inset. Instead, the projector quickly and abruptly converges to a fixed position as we decrease $\rho$. Attempting to extract the localization length via the second moment, as shown in the right panel, we obtain a convergence to an $L$-independent constant in agreement with Fig.~\ref{fig:method1_Haldane}(c). This demonstrates the difference between topological and trivial phases at the level of the projector. Just as localization length divergence in topological phases corresponds to self-similar projector expansion, the convergence of the localization length in trivial phases corresponds to projector convergence in real space.

%%%%%%%%%%%%%%%%%%%%%%
\section{Discussion} %
\label{sec:disc}     %
%%%%%%%%%%%%%%%%%%%%%%

Following these results, in this section we discuss the scaling exponent. In particular, we comment on the RG framework of the truncation procedure in Sec.~\ref{subsec:rg}, its analogy with the plateau transition in Sec.~\ref{subsec:recover}, and its utility in diagnosing topological phases in Sec.~\ref{subsec:diagnose}.

\subsection{Real-space RG framework}
\label{subsec:rg}

As mentioned in Sec.~\ref{sec:loc_norm}, in order for localization renormalization to hold, it is a necessary condition that we can construct a renormalized system $\{H', \ket{\psi'}\}$ from the original system $\{H, \ket{\psi}\}$. In the single-particle case, one way of demonstrating this is to show that the family of projectors $P_\rho$, defined in Eq.~\eqref{eq:sp_proj}, is self-similar, which implies that the truncated system can be related to the original system under a coarse-graining, rescaling, and renormalization. By analyzing the expansion of real-space projectors on each $\rho$ iteration, we have demonstrated that we can consolidate our state removal procedure into such an RG framework~\cite{Kadanoff66}. Each step of removing the most-localized states in the system is analogous to coarse-graining the Hamiltonian to an effective system of remaining states $\bar{H}(\mathbf{r})=P_\rho H(\mathbf{r})P_\rho$. Subsequently, we have shown, by studying the projectors in Figs.~\ref{fig:method1_nLL_proj} and~\ref{fig:method1_Haldane_proj}, that the system is statistically self-similar under a rescaling $\mathbf{r}'=\rho^{-1/2}\mathbf{r}$ and renormalization $H'(\mathbf{r}')=\rho^{-1}\bar{H}(\mathbf{r}')$. Although the projectors do not exhibit an exact self-similarity in our examples, they are statistically self-similar to a high degree --- approximately up to the tenth moment in the configurations that we study. Moreover, since the localization length is conventionally defined via the second moment of displacement, it can be viewed as a self-similar property of the family of projectors $P_\rho$ and localization renormalization may be applied.

\subsection{Quantum Hall plateau transition}
\label{subsec:recover}

The most famous example of a localization transition induced by the topology-localization dichotomy in quantum Hall systems is the plateau transition, and so it is natural to ask whether an analogy can be made with the localization scaling studied in this paper. Although these two concepts are fundamentally different, with the plateau transition yielding a universal scaling exponent of $\nu_\text{p}\simeq2.5$~\cite{wei88, li09, chalker88, huckestein92, huo92, slevin09, zhu19}, there are certain qualitative comparisons that can be drawn. In localization renormalization, we induce a phase transition by explicitly projecting out the most-localized states at individual points in space. On the other hand, in the plateau transition, we induce a phase transition by tuning the Fermi energy with respect to a disorder potential landscape. Loosely speaking, there is a correspondence between maximally-localized states and states that are trapped around extrema in a disorder landscape. In this language, localization renormalization equates to trapping isolated states using equal-amplitude Dirac delta extrema in a fictitious potential, whereas the plateau transition equates to trapping eigenstates of a real disorder potential $V(\mathbf{r})\approx \sum_i V_i \delta(\mathbf{r}-\mathbf{r}_i)$, where $V_i$ is a random amplitude at position $\mathbf{r}_i$. Although localization renormalization is a numerical algorithm and does not correspond to a physically-motivated phase transition, it would be interesting to explore this analogy in future work~\cite{Liu23, Mildner23}.

%In particular, the scaling corresponding to plateau jumps in the integer quantum Hall effect has a starkly different universal exponent, widely-believed to be $\nu_\text{p}\simeq2.5$~\cite{wei88, li09, chalker88, huckestein92, huo92, slevin09, zhu19}. In order to reconcile these two results, we consider a system described by a generic quantum Hall Hamiltonian $H_\text{QH}=T_{\text{LL}}+V$, where $T_\text{LL}$ is the Landau level kinetic energy contribution and $V$ is a disorder component. As disorder is introduced in the system, the degeneracy of the basis states is lifted and an orthogonal basis is selected for the Hilbert space, with varying degrees of localization. The main difference arises since we project out the most-localized eigenstates of individual extrema projected to the LLL (or lowest band) $V=V_0 P_\text{LLL} \delta(\mathbf{r}) P_\text{LLL}$, whereas in a realistic model of disorder, we would instead project out eigenstates of the effective disorder component $V\approx \sum_i V_i \delta(\mathbf{r}-\mathbf{r}_i)$, where $V_i$ is a random amplitude at position $\mathbf{r}_i$. We note that localization renormalization is a numerical algorithm and does not correspond to a physically-motivated phase transition. However, by repeating the computations to eliminate states with respect to a random disorder component, we can recover the universal scaling exponent $\nu_\text{p}$.

\subsection{Topological phase diagnosis}
\label{subsec:diagnose}

The broad scope of the localization renormalization procedure is to classify a variety of condensed matter phases using universal scaling exponents. Based on our results, we conjecture that $\nu\sim1/2$ for all two-dimensional class A topological insulators, and we generally expect different scaling exponents in other symmetry classes and dimensions, e.g.~$\nu\sim 1/d$. However, although the procedure is designed as a numerical tool to study a diverse selection of localization transitions, the examples shown in this paper are all centered on the topology-localization dichotomy, and so the method doubles as a technique for diagnosing topological and trivial phases. This has a number of advantages compared to conventional approaches, such as computing edge modes or the Chern number. For example, the algorithm is spectrum independent and so may be used to diagnose topology in disordered systems, provided the disorder is sufficiently weak so as to not mix the bands. Moreover, the method is numerically inexpensive, with a clear power-law divergence and scaling exponent reported after only the first few removal iterations\footnote{Provided that the parameter set is sufficiently far from a phase transition.}. These advantages can prove particularly useful in cases where the phase diagram is unknown or traditional methods for computing the Chern number fail, such as in fractal lattices~\cite{Jha23}, hyperbolic lattices~\cite{Zhang23}, and quasicrystals~\cite{Koshino22}, as well as in systems with other symmetry classes or higher dimensions.

%%%%%%%%%%%%%%%%%%%%%%
\section{Conclusion} %
\label{sec:conc}     %
%%%%%%%%%%%%%%%%%%%%%%

We have introduced the localization renormalization formalism as a way of analyzing a diverse range of localization transitions, which we demonstrated numerically using integer quantum Hall examples. By iteratively removing an orthogonal subset of maximally-localized states from distinct sites, we can induce quantum Hall breakdown transitions with a localization length divergence of $\lim_{\rho\to 0, L\to\infty}\xi(\rho) \sim \rho^{-\nu}$ with $\nu=0.5$ in topological systems, and convergence to a system-size-independent constant in the trivial case. The scaling exponent in these topological systems is universal, and therefore independent of the model used to describe the system, the metric used to define the localization length, and the details of the state removal algorithm. Moreover, by analyzing the expansion of real-space projectors on each iteration, we find that the scaling exponent is a self-similar property of the family of projectors $P_\rho$, which accords with an RG picture. This motivates the use of localization renormalization as a versatile diagnostic tool for quantum Hall systems, in cases where traditional methods of diagnosing band topology are inadequate, as well as in topologically trivial systems and beyond.

\begin{acknowledgments}
We thank Glenn Wagner, Samuel Garratt, Michael Zaletel, Adrian Culver, Pratik Sathe, David Bauer, Albert Brown, Xu Liu, and Spenser Talkington for useful discussions. In particular, we thank Glenn Wagner and Samuel Garratt for helpful comments on the manuscript. Preliminary results from this work are presented in the Ph.D.~thesis of Dominic Reiss~\cite{reiss20}. B.A.~acknowledges support from the Swiss National Science Foundation under grant no.~P500PT\_203168; B.A.~and R.R.~acknowledge support from the University of California Laboratory Fees Research Program funded by the UC Office of the President (UCOP), grant ID~LFR-20-653926; D.R., F.H., and R.R.~acknowledge support from the NSF under CAREER grant no.~DMR-1455368 and the Alfred P.~Sloan foundation; and all authors thank the Mani L.~Bhaumik Institute for Theoretical Physics.
\end{acknowledgments}

\bibliographystyle{apsrev4-1}
\bibliography{ms}

\end{document}

%%%%%%%%%%%%%%%%%%%%%%%%%%%%%%%%%%%%%%%%%%%%%%%%%%%%%%%%%%%%%%%%%%%%%%%%%%%%%%%%%%%%%%%%%%%%%%%
%%% SUPPLEMENTARY MATERIAL %%%%%%%%%%%%%%%%%%%%%%%%%%%%%%%%%%%%%%%%%%%%%%%%%%%%%%%%%%%%%%%%%%%%%%%%
%%%%%%%%%%%%%%%%%%%%%%%%%%%%%%%%%%%%%%%%%%%%%%%%%%%%%%%%%%%%%%%%%%%%%%%%%%%%%%%%%%%%%%%%%%%%%%%

\onecolumngrid

%%%%%%%%%% Merge with supplemental materials %%%%%%%%%%

\begin{center}
	\textbf{\large Supplementary Material: Localization renormalization and quantum Hall systems}
\end{center}

\tableofcontents

%%%%%%%%%% Merge with supplemental materials %%%%%%%%%%
%%%%%%%%%% Prefix a "S" to all equations, figures, tables and reset the counter %%%%%%%%%%
\setcounter{equation}{0}
\setcounter{figure}{0}
\setcounter{table}{0}
\setcounter{section}{0}
\makeatletter
\renewcommand{\theequation}{S\arabic{equation}}
\renewcommand{\thefigure}{S\arabic{figure}}
\renewcommand{\thetable}{S\arabic{table}}
\renewcommand{\thesection}{S\Roman{section}}
\renewcommand{\bibnumfmt}[1]{[S#1]}
\renewcommand{\citenumfont}[1]{S#1}
%%%%%%%%%% Prefix a "S" to all equations, figures, tables and reset the counter %%%%%%%%%%

\section{Symmetric orthogonalization}
\label{sec:lowdin}

In this section, we describe the symmetric orthogonalization procedure, also known as L{\"o}wdin orthogonalization~\cite{lowdin70}.

Given a set of linearly independent, but not necessarily orthonormal, vectors $\{\ket{\psi}\}$, we form the operator $S$ whose matrix elements (indexed by the same set that indexes the input vectors $\{\ket{\psi}\}$) are given by the overlaps $S_{ij} = \braket{\psi_i|\psi_j}$. The operator $S$ is Hermitian and is known as the Gram matrix of the spanning set $\{\ket{\psi}\}$. The eigenvalues of Gram matrices are non-negative by construction. Furthermore, since we posit that the set $\{\ket{\psi}\}$ is linearly independent, we can show that the eigenvalues of $S$ are positive. We can then form the orthogonal set $\{\ket{\tilde{\psi}}\}$ as
\begin{equation}
\ket{\tilde{\psi}_j} = \sum_i S^{-1/2}_{ij} \ket{\psi_i}.
\end{equation}

As opposed to iterative methods, such as Gram-Schmidt orthogonalization, the symmetric orthogonalization treats each input vector on an equal footing. In fact, it can be shown that when using the symmetric orthogonalization, the resulting functions $\ket{\tilde{\psi}_j}$ are closest to the input vectors in a least-squares sense, that is the quantity $\sum_j || \ket{\tilde{\psi}_j} - \ket{\psi_j} ||^2$ is minimized~\cite{carlson57,mayer02}.

\section{Alternative methods to truncate the single-particle basis}

In this section, we present two alternative procedures for Hilbert space truncation, based on the simultaneous removal of states centered on sublattices in Sec.~\ref{sec:sublattice_elim}, and the sequential removal of individual states in Sec.~\ref{sec:single_elim}. In both cases, we find that the localization scaling exponent $\nu$ agrees with the main text.

\subsection{Sequential elimination of sublattices of maximally-localized states}
\label{sec:sublattice_elim}

\subsubsection{Method}
\label{subsec:sublattice_procedure}

\begin{figure}
	\centering
	\includegraphics[width=0.3\linewidth]{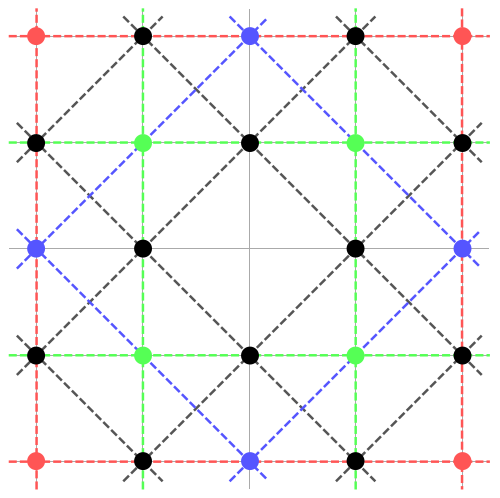}
	\caption{\label{fig:gradedsublattices} \textbf{Lattice decomposition.} An illustration of the decomposition of the underlying lattice (less the origin) $\mathcal{L} = \bigoplus_{n=0} \mathcal{L}_n$, centered at the origin. $\mathcal{L}_0$, $\mathcal{L}_1$, $\mathcal{L}_2$, and $\mathcal{L}_3$ are depicted in black, green, blue, and red, respectively. Each subsequent sublattice $\mathcal{L}_i$ has growing lattice constant $a_n = \sqrt{2} a_{n-1}$ and is rotated by $\pi/4$ with respect to the previous sublattice.}
\end{figure}

For this procedure, we introduce a square lattice $\mathcal{L}$ of lattice spacing $a = \sqrt{A}$, where $A$ is the real-space area occupied per state. In Landau levels, we introduce $\mathcal{L}$ such that the origin of the lattice lies at the origin of the plane, whereas in the Haldane model, we simply take $\mathcal{L}$ to be the lattice in the real-space definition of the model. Moreover, we introduce a decomposition of the lattice (less the origin) $\mathcal{L} = \bigoplus_{n = 0} \mathcal{L}_n$. Each subsequent sublattice in the decomposition $\mathcal{L}_n$ has growing lattice spacing $a_n = \sqrt{2^{n+1}}a$, and is translated and rotated such that it is disjoint from the other sublattices and the origin. The sublattices $\mathcal{L}_n$ are illustrated in Fig.~\ref{fig:gradedsublattices}.

This construction decomposes $\mathcal{H}$ as a direct sum $\mathcal{H} = \bigoplus_{n=0} \mathcal{H}_n$, where each subspace $\mathcal{H}_n$ has a mutually orthogonal basis of states $\{\ket{\tilde{\psi}_{n}}\}$ and each state is localized in real space about points $\mathbf{r}_{ij}$ in the $n$th sublattice $\mathcal{L}_n$. Each subsequent Hilbert space is therefore approximately half the dimension of its predecessor, i.e.~$\dim(\mathcal{H}_0) \simeq N/2$, $\dim(\mathcal{H}_1) \simeq N/4$, etc. Given a projection operator $P = P_0$ onto the occupied states, we now detail an iterative procedure for finding and eliminating the basis $\{\ket{\tilde{\psi}_{n}}\}$. Beginning with $n = 0$:

\begin{enumerate}
	\item For each site $\mathbf{r}_{ij}$ in the $n$th sublattice $\mathcal{L}_n$, we define the state $\ket{\psi_{n,ij}}$ as the eigenvector corresponding to the minimum non-zero eigenvalue of the projected distance-squared operator $D_{n,ij}^2 = P_n(\mathbf{r}-\mathbf{r}_{ij})^2P_n$. 
	\item We perform a symmetric orthogonalization of the states $\{\ket{\psi_{n}}\} \rightarrow \{\ket{\tilde{\psi}_{n}}\}$ following the procedure in Sec.~\ref{sec:lowdin}.
	\item We remove the selected states from the projection operator for the next iteration, that is $P_{n+1} = P_n - \sum_{i,j\in\mathcal{L}_n} \ket{\tilde{\psi}_{n,ij}}\bra{\tilde{\psi}_{n,ij}}$. If the rank of $P_{n+1}$ is not zero, then we perform another iteration with $n \rightarrow n+1$. Otherwise, we have exhausted $\mathcal{H}$. 
\end{enumerate}

After each iteration of removing a sublattice of localized degrees of freedom, we analyze the remaining localized degrees of freedom. We associate to the projector $P_{n}$ the density parameter $\rho_n = 2^{-n}$, which represents the fraction of the original Hilbert space dimension supported by $P_{n}$. 

The origin of $\mathcal{L}$, due to the symmetry of the procedure, is the most isolated lattice point at any given step. We therefore expect that from the remaining degrees of freedom, we may construct a maximally-localized degree of freedom about the origin\footnote{At each step there will be points in the lattice that are equivalent to the origin, and one could choose any of these to find the same localization length. In our case, it is best to select the origin as it will be least effected by errors accrued near the boundaries.}. At the $n$th step, we find the eigenstate $\ket{\mathbf{0}_{\rho_n}}$ satisfying $P_n \mathbf{r}^2 P_n = \xi^2(\rho_n)\ket{\mathbf{0}_{\rho_n}}$, where, in analogy with the main text, $\xi^2(\rho_n)$ is the smallest non-zero eigenvalue of the projected distance-squared operator.

\subsubsection{Results}

\paragraph{LLL}

We choose the lattice $\mathcal{L}$ to be square with lattice vectors aligned with $\hat{\mathbf{e}}_x$ and $\hat{\mathbf{e}}_y$, such that the origin of the plane is coincident with the origin of the lattice.

\begin{figure}
	\centering
	(a)~LLL\\
	\vspace{1em}
	\includegraphics[width=.5\linewidth]{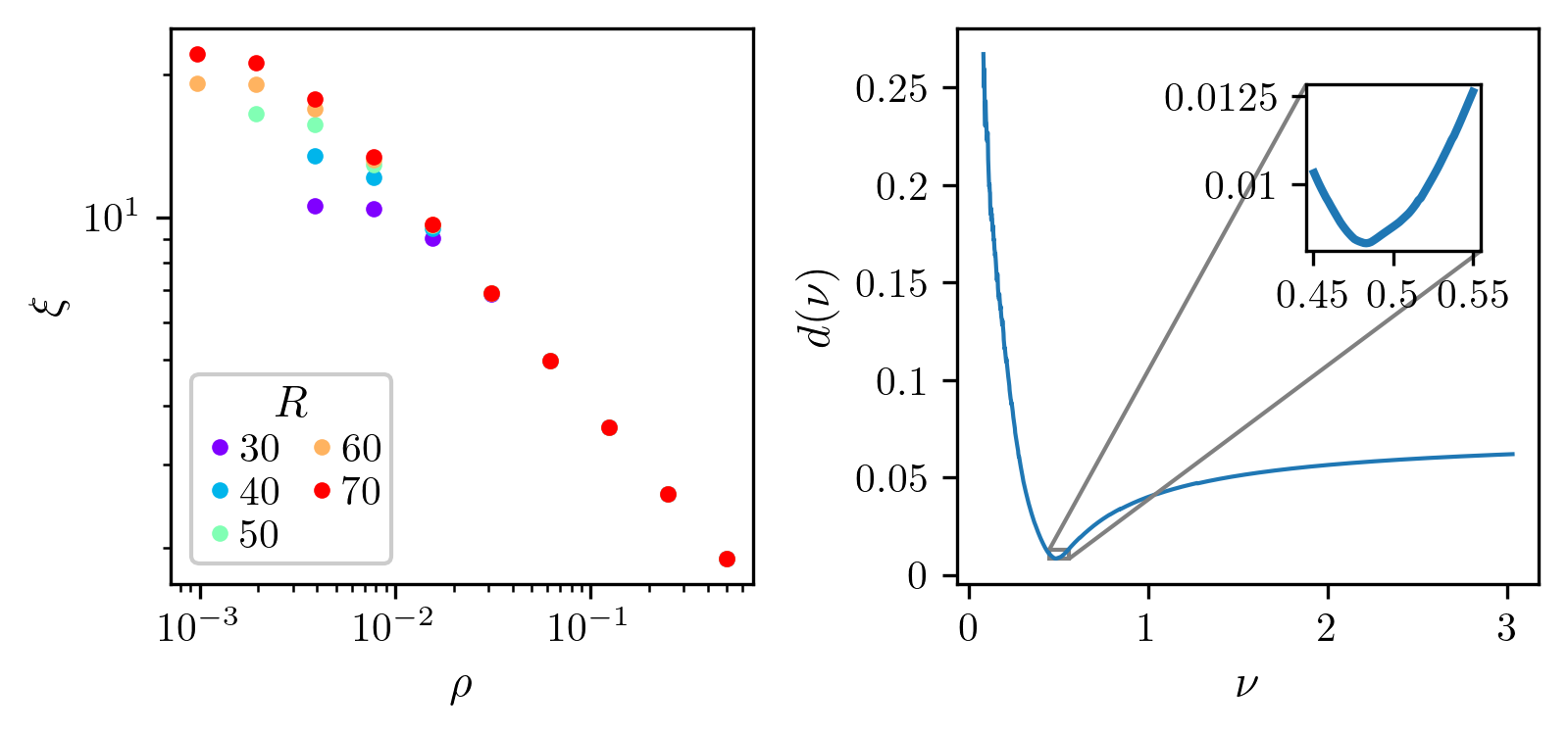}\\
	(b)~Haldane model\\
	\vspace{1em}
	\includegraphics[width=.5\linewidth]{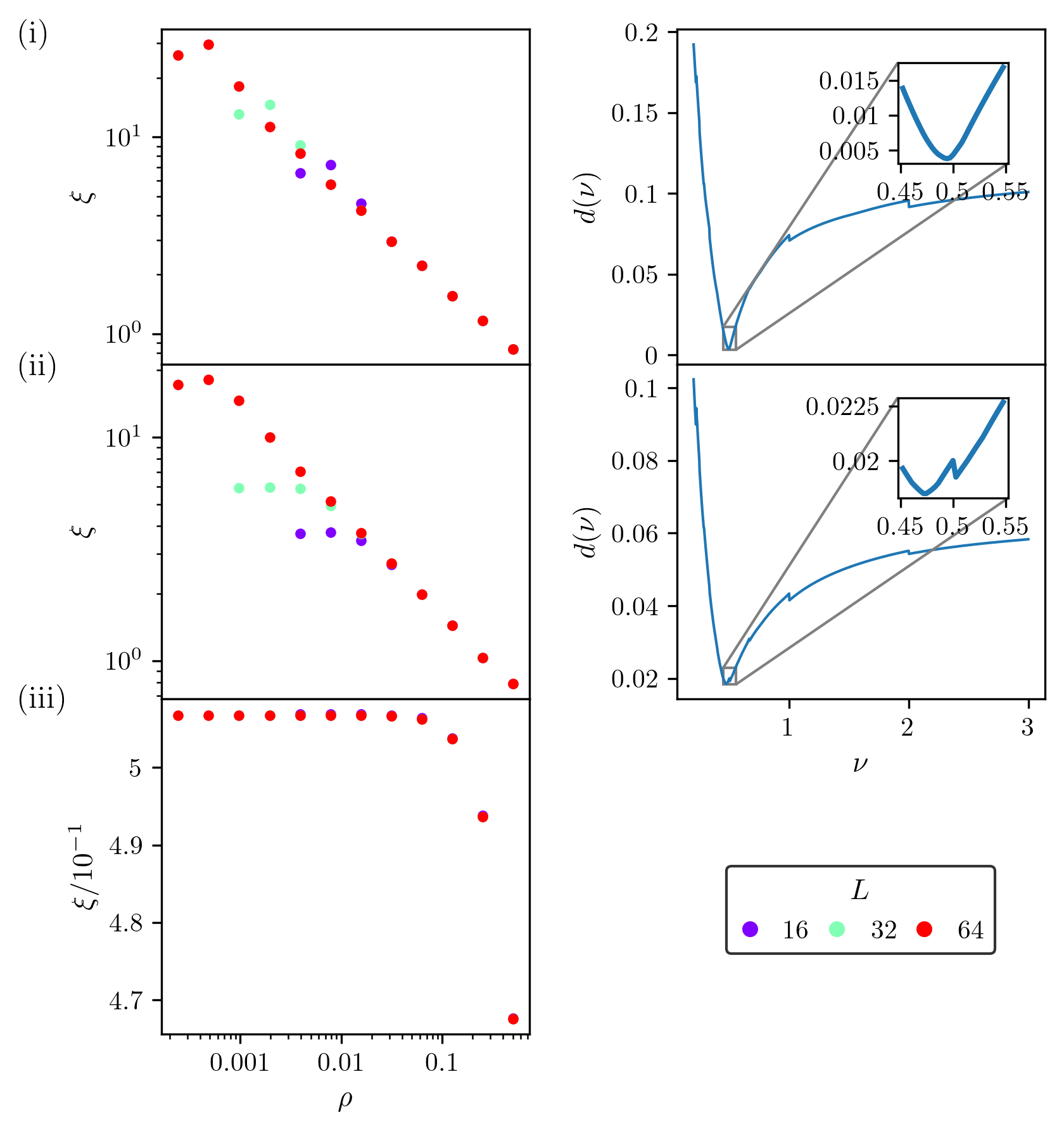}
	\caption{\label{fig:method2} \textbf{Sequential elimination of sublattices of maximally-localized states.} (left)~Localization length $\xi(\rho)$ obtained for (a)~the LLL, and (b)~the Haldane model with (i)~$\mathbf{t}_a$, (ii)~$\mathbf{t}_b$, and (iii)~$\mathbf{t}_c$, using the sublattice elimination procedure described in Sec.~\ref{subsec:sublattice_procedure}. The system size $L$ is depicted in different colors. (right)~The quality function of data collapse $d(\nu)$ for the data depicted on the left. Optimization of the quality function results in critical exponents of (a)~$\nu_\text{LLL} = 0.483 \pm 0.036$, (b)(i)~$\nu_{\mathbf{t}_a} = 0.495 \pm 0.010$, and (b)(ii)~$\nu_{\mathbf{t}_b} = 0.472 \pm 0.049$.}
\end{figure}

We perform the procedure for various domain radii $R$. In particular, the domains considered support between $N = 450$ and $2450$ states. For each initial dimension $N$, we perform iterations until the Hilbert space is exhausted. Using this procedure, each system size can now access only the discrete set of values $\rho_n = 2^{-n}$, and a system of dimension $N$ can access up to $n \simeq \log_2(N)$. The length $\xi(\rho)$ for various $N$ is plotted in the left panel of Fig.~\ref{fig:method2}(a) and diverges with a truncated power law, as expected.

We again perform optimization of the collapse quality function $d(\nu)$ with respect to the localization length exponent $\nu$. The right panel of Fig.~\ref{fig:method2}(a) depicts the quality function of collapse over a range of $\nu$, again indicating optimal $\nu \simeq 0.5$. Optimization was performed using the Nelder-Mead descent method determining $\nu_\text{LLL} = 0.483 \pm 0.036$.

\paragraph{Haldane model}

When applying the sublattice elimination procedure to the Haldane model, we limit ourselves to system sizes  $N=L \times L$, where $L$ is a power of two, in our case $L = 16, 32, 64$ (or equivalently, $N = 256$, $1024$, $4096$). If $L$ is not a power of two, each subsequent sublattice $\mathcal{L}_{n+1}$ in the decomposition will eventually fail to comprise half of the states of $\mathcal{L}_n$. Moreover, the symmetry of the origin site will be broken at this point, potentially invalidating our assumption that after any given iteration a maximally-localized remaining degree of freedom can be found there.

The $\xi(\rho)$ data for the parameters within the topological phase, $\mathbf{t}_a$ and $\mathbf{t}_b$, are shown in the left panels of Figs.~\ref{fig:method2}(b)(i,ii). Again we see that in the topological phase, the localization lengths $\xi$ diverge as a power law. Optimization of data collapse, depicted in the right panels of Figs.~\ref{fig:method2}(b)(i,ii), results in $\nu_{\mathbf{t}_a} = 0.495 \pm 0.010$ and $\nu_{\mathbf{t}_b} = 0.472 \pm 0.049$.

The effective localization lengths $\xi(\rho)$ for the Haldane model at $\mathbf{t}_c$ within the trivial phase are shown in Fig.~\ref{fig:method2}(b)(iii). The length $\xi(\rho)$ quickly saturates to an $N$-independent value and does not exhibit a power-law divergence as $\rho \rightarrow 0$.

\subsubsection{Discussion}
\label{subsec:sublattice_disc}

The data from this procedure closely resemble that presented in the main text. As expected, the trivial system shows no singular behavior in $\xi$, and the topological systems yield $\nu \simeq 0.5$ within their respective uncertainty intervals.

It may be intuitively deduced that the elimination of states on expanding sublattices described in this section should produce the same exponent $\nu$ as in the main text. Each iteration of this sublattice elimination procedure is equivalent to the simultaneous elimination procedure with $\rho_1 = 1/2$, and each subsequent iteration can be interpreted as performing the simultaneous elimination procedure but replacing the `original/underlying' system after each iteration with the current effective system. The fact that these two methods are quantitatively consistent suggests that the exponent $\nu \simeq 0.5$ is universal across all intermediate effective systems derived using $P_\rho$.

The sublattice structure intrinsic to this elimination procedure provides a natural picture of how the localization properties of each effective system evolve as $\rho \rightarrow 0$. In our original system, each state occupies a real-space area $A$. When we eliminate a localized degree of freedom centered at some $\mathbf{r}$, we can consider that the effective Hilbert space no longer supports a localized function centered within an area $A$ around $\mathbf{r}$. After removing half of the degrees of freedom, the distances between potential centers for localized degrees of freedom increase by a factor $\sqrt{2}$, precisely the growth of $\xi$ that we observed. This suggests that removing the most-localized degrees of freedom screens the remaining effective localized degrees of freedom, so that their decay is similar to the states of the original system in some scaled coordinates $\mathbf{r}^\prime = \sqrt{2}\mathbf{r}$.

\subsection{Sequential elimination of individual maximally-localized states}
\label{sec:single_elim}

\subsubsection{Method}
\label{subsec:single_procedure}

As in Sec.~\ref{sec:sublattice_elim}, we introduce a square lattice $\mathcal{L}$ of lattice spacing $a = \sqrt{A}$, where $A$ is the real-space area occupied per state. 

Given a projection operator $P = P_0$ onto the space of occupied states (of dimension $N$), we now detail a procedure to iteratively eliminate the single most-localized state. Since we are now eliminating one state at a time, after step $n$ the fraction of remaining states $\rho_n$ is $\rho_n = 1 - n/N$. Beginning with $n=0$:

\begin{enumerate}
	\item For each site $\mathbf{r}_{ij}$ in $\mathcal{L}$, we define the state $\ket{\psi_{n,ij}}$ as the eigenvector corresponding to the minimum non-zero eigenvalue $\xi^2_{ij}(\rho_n)$ of the projected distance-squared operator $D_{n,ij}^2 = P_n(\mathbf{r}-\mathbf{r}_{ij})^2P_n$. 
	\item We select the state $\ket{\psi_{n,IJ}}$ corresponding to the site with the minimum second moment, that is    
	\begin{equation}
	\label{eq:single_procedure_loclen}
	\xi^2_{IJ}(\rho_n) = \min_{ij}\left\{\xi^2_{ij}(\rho_n)\right\}.
	\end{equation}
	If there is a degenerate minimum, we freely choose any such $IJ$.
	\item We remove the selected state from the projection operator for the next iteration, that is $P_{n+1} = P_n - \ket{\psi_{n,IJ}}\bra{\psi_{n,IJ}}$. If the rank of $P_{n+1}$ is not zero, then we perform another iteration with $n \rightarrow n+1$. Otherwise, we have exhausted $\mathcal{H}$. 
\end{enumerate}

In contrast to the previous procedures, which were symmetric about the origin, this procedure does not guarantee symmetry. At each step $n$, we can associate the minimum second central moment (as defined in Eq.~\eqref{eq:single_procedure_loclen}) $\xi(\rho_n)$ $(= \xi_{IJ}(\rho_n))$ with the density $\rho_n$.

\subsubsection{Results}

\paragraph{LLL}

We perform the procedure for various circular domains, comprising systems of dimension between $N = 200$ and $900$ states. These system sizes are smaller than the ones used for the procedures discussed in the main text and Sec.~\ref{sec:sublattice_elim} because this iterative process is more computationally expensive.

\begin{figure}
	\centering
	(a) LLL\\
	\vspace{1em}
	\includegraphics[width=.5\linewidth]{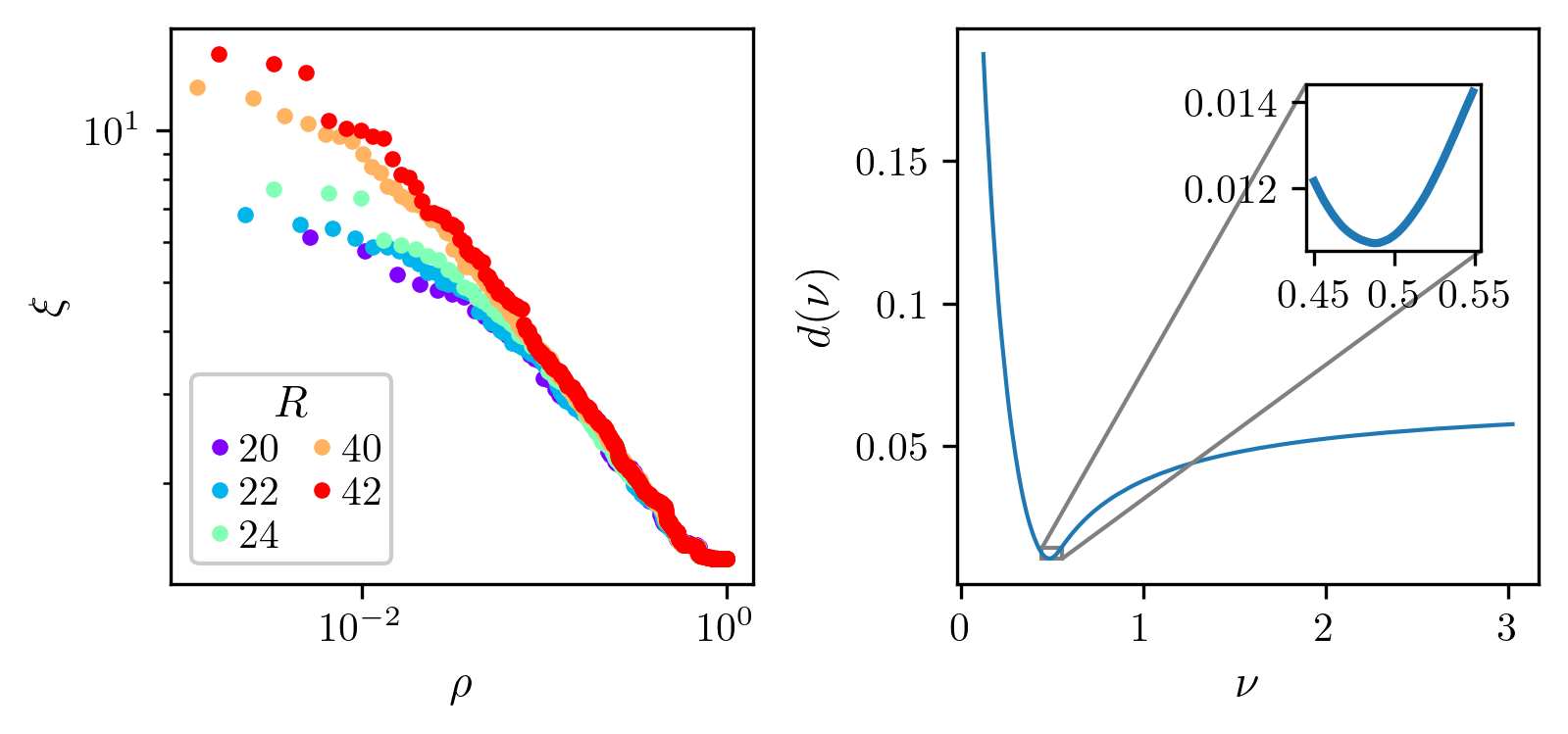}\\
	(b) Haldane model\\
	\vspace{1em}
	\includegraphics[width=.5\linewidth]{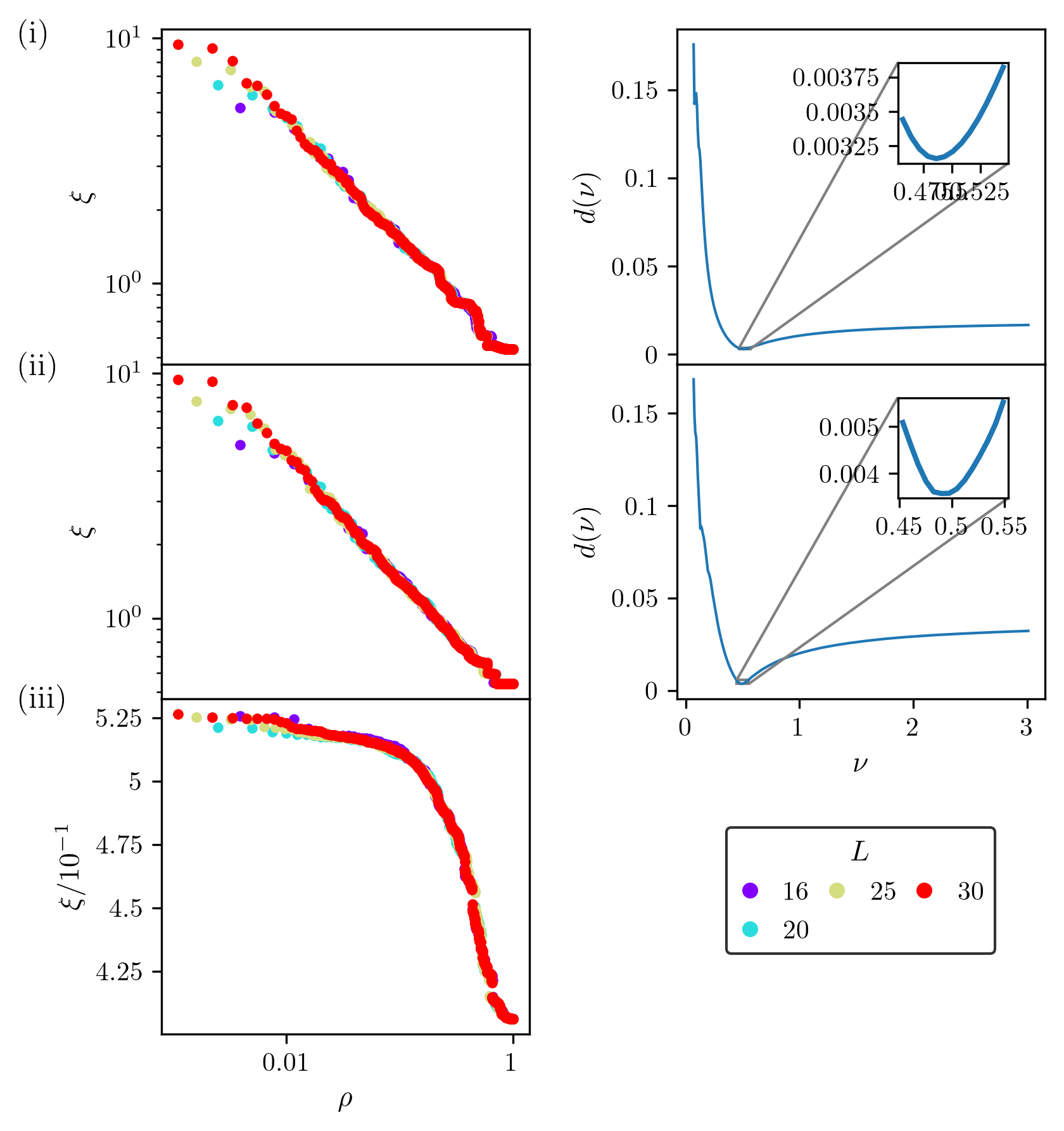}
	\caption{\label{fig:method3} \textbf{Sequential elimination of individual maximally-localized states.} (left)~Localization length $\xi(\rho)$ obtained for (a)~the LLL, and (b)~the Haldane model with (i)~$\mathbf{t}_a$, (ii)~$\mathbf{t}_b$, and (iii)~$\mathbf{t}_c$, using the state elimination procedure described in Sec.~\ref{subsec:single_procedure}. The system size $L$ is depicted in different colors. (right)~The quality function of data collapse $d(\nu)$ for the data depicted on the left. Optimization of the quality function results in critical exponents of (a)~$\nu_\text{LLL} = 0.487 \pm 0.049$, (b)(i)~$\nu_{\mathbf{t}_a} = 0.483 \pm 0.069$, and (b)(ii)~$\nu_{\mathbf{t}_b} = 0.478 \pm 0.045$.}
\end{figure}

For each system dimension $N$, we iteratively eliminate the state with the minimum localization length from the system. It is numerically infeasible to perform this minimization in the space of normalized functions. Instead we introduce a lattice $\mathcal{L}$ of unit cell area $2\pi$ of candidate function centers, and find the maximally-localized function with respect to each of these centers. We then approximate the global minimum to be the minimum of functions centered on this lattice. The minimal localization lengths $\xi(\rho)$ for various $N$ are plotted in the left panel of Fig.~\ref{fig:method3}(a).

We again perform optimization of $d(\nu)$ with respect to the localization length exponent $\nu$. The right panel of Fig.~\ref{fig:method3}(a) depicts the quality function of collapse over a range of $\nu$, again indicating optimal $\nu \simeq 0.5$. Optimization was performed using the Nelder-Mead descent method determining $\nu_\text{LLL} = 0.487 \pm 0.049$.

\paragraph{Haldane model}

We perform the single-state elimination procedure for square systems with periodic boundary conditions, of dimension $N = L \times L$ between $L = 16$ and $30$ (or equivalently, between $N = 256$ and $900$).

The $\xi(\rho)$ data for the parameters within the topological phase, $\mathbf{t}_a$ and $\mathbf{t}_b$, are shown in the left panels of Figs.~\ref{fig:method3}(b)(i,ii). Again we see that in the topological phase, the localization lengths $\xi$ diverge as a power law. Optimization of data collapse, depicted in the right panels of Figs.~\ref{fig:method3}(b)(i,ii), results in $\nu_{\mathbf{t}_a} = 0.483 \pm 0.069$ and $\nu_{\mathbf{t}_b} = 0.478 \pm 0.045$.

The effective localization lengths $\xi(\rho)$ for the Haldane model at $\mathbf{t}_c$ within the trivial phase are shown in Fig.~\ref{fig:method3}(b)(iii). Within the trivial phase, $\xi(\rho)$ again quickly saturates to an $N$-independent value and, as expected, does not exhibit a power-law divergence as $\rho \rightarrow 0$.

\subsubsection{Discussion}

As before, the single-state removal procedure results in $\nu \simeq 0.5$ for each topological system we considered. Na\"ively, this procedure may appear closer to emulating a plateau transition, as compared to the other more symmetric procedures. However, in the thermodynamic limit, removing a single state iteratively is not well defined. If we instead interpret this method as removing a fraction $1/N$ of the maximally-localized states, we see that each iteration of this method is equivalent to the simultaneous elimination procedures. It is therefore unsurprising that we again find $\nu$ consistent with the previous methods.

\section{Alternative removal lattices for the truncation algorithm}
\label{sec:non-square}

\begin{figure}
	\centering
	\includegraphics[width=.5\linewidth]{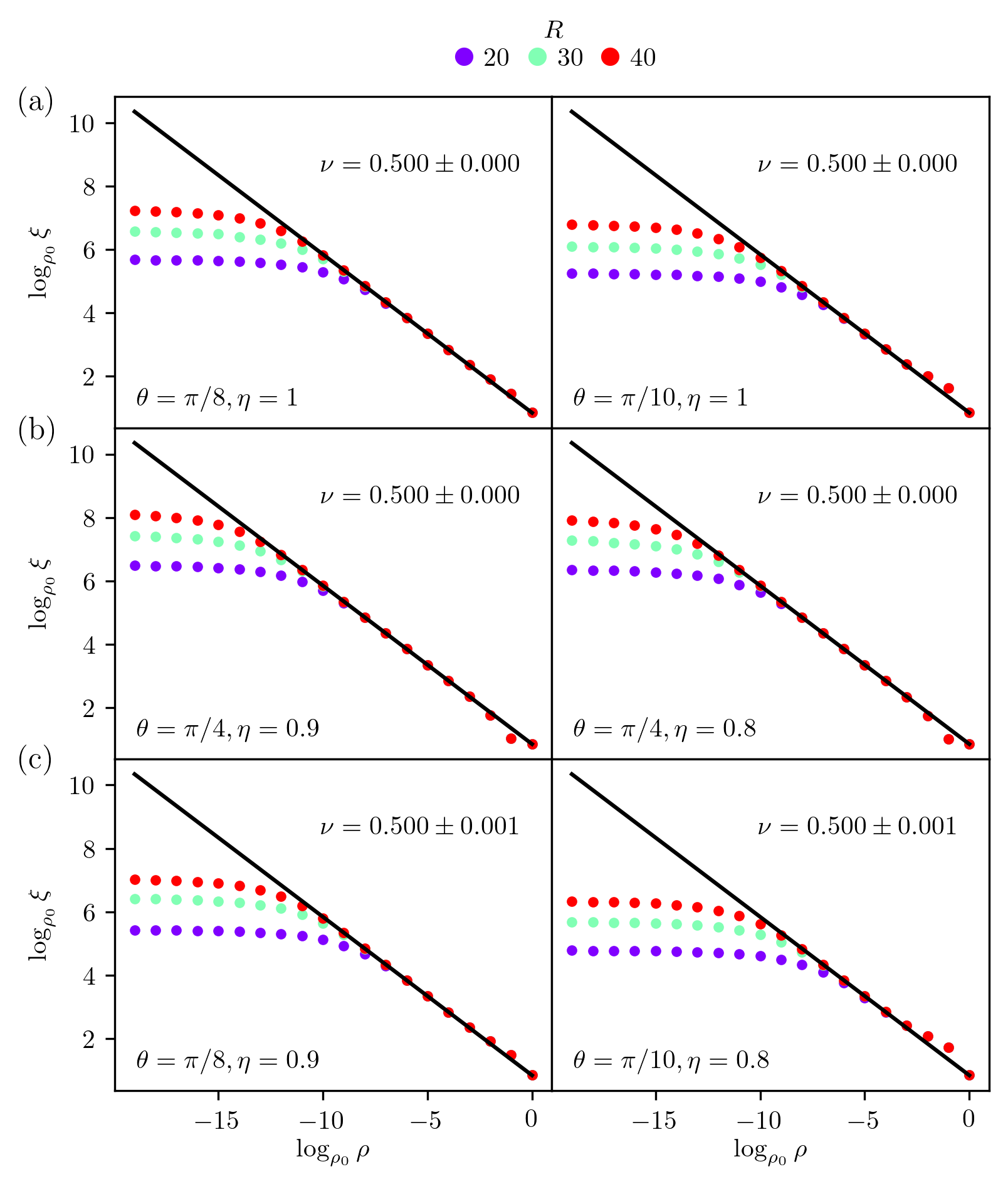}
	\caption{\label{fig:non-square} \textbf{Scaling for non-square removal lattices.} Localization length $\xi(\rho)$ obtained for the LLL, with $\rho=\rho_0^{-\kappa}$, $\kappa\in\mathbb{N}$, and $\rho_0=1.5$, using a parallelogram lattice to remove states. The scaling is shown for (a)~oblique lattices with an opening angle $2\theta$, (b)~anisotropic lattices with an anisotropy parameter $\eta=a_2/a_1$, and (c)~a combination of the two. The results are also presented with respect to system size $R$, and the line of best fit is drawn for $R=40$, through the linear region.}
\end{figure}

In the main text, we stated that we can choose a square lattice to remove localized states without loss of generality. In this section, we explicitly show that the scaling relation and scaling exponent $\nu\simeq0.5$ does not depend on the geometry of the removal lattice $\mathcal{L}_{\rho}$. To this end, we consider, as an example, the scaling for the LLL corresponding to the square lattice in Fig.~2(a) of the main text. We proceed to perform the Hilbert space truncation procedure for this system, described in the main text, using a generic parallelogram lattice. In analogy to the square lattice with basis vectors $\mathbf{a}_1=a\hat{\mathbf{e}}_x, \mathbf{a}_2=a\hat{\mathbf{e}}_y$ satisfying $\mathbf{a}_1\cdot\mathbf{a}_2=0$, the parallelogram lattice has basis vectors $\mathbf{a}_1=a\hat{\mathbf{e}}_x, \mathbf{a}_2=\eta a (\cos(2\theta) \hat{\mathbf{e}}_x+\sin(2\theta)\hat{\mathbf{e}}_y)$ satisfying $\mathbf{a}_1\cdot\mathbf{a}_2=\eta a^2 \cos(2\theta)$, where $\eta$ is the anisotropy parameter. Hence, the obliqueness of the lattice is quantified by the opening angle $2\theta$ and the anisotropy is given by the ratio of basis vector lengths. In the case of a square lattice, we removed states on a disc of radius $R$, with coordinates satisfying $x^2+y^2<R^2$, and so analogously we now remove states on an ellipse, with coordinates satisfying $\tilde{x}^2+(\tilde{y}/\eta)^2<R^2$. Moreover, in order to preserve the unit cell area of $2\pi$, the lattice constant now takes the general form $a_\rho=\sqrt{2\pi/((1-\rho)\eta\sin(2\theta))}$. In Fig.~\ref{fig:non-square}, we show the scaling corresponding to Fig.~2(a) using a parallelogram lattice with various values for the obliqueness and anisotropy. These results show that, even with modest system sizes, the scaling relation is recovered in each case and so the exponent does not depend on lattice geometry. We note, however, that for highly distorted lattices (cf.~the right panel of Fig.~\ref{fig:non-square}(c)), the scaling relation does not hold as well for large values of $\rho$, and so it is advantageous to use a square lattice in numerical simulations.

\section{Alternative methods to compute the localization length}
\label{sec:alt_nu}

In addition to the three methods studied to truncate the Hilbert space, detailed in the main text, Sec.~\ref{sec:sublattice_elim}, and Sec.~\ref{sec:single_elim}, there are also a number of ways to compute the scaling exponent $\nu$. Throughout this paper, we have computed the localization length by defining $\xi^2$ as the minimum eigenvalue of the $P_\rho D^2 P_\rho$ matrix. However, we may equivalently define the localization length via a matrix element, for example via $\xi^2=\braket{\mathbf{0}|P_\rho D^2 P_\rho|\mathbf{0}}$, where $\ket{\mathbf{0}}$ denotes the origin. Alternatively, as we saw in Secs.~III~A~3 and~III~B~3 of the main text, we may extract the localization length from the moments of the projector, such that $\xi^\gamma = \braket{r^\gamma}$, where $r$ is the radius, and the expectation value is taken with respect to the projector probability distribution. All of these methods are consistent and yield the same value for the scaling exponent $\nu$ in the $\rho\to0$ and $N\to\infty$ limits.

In the case of higher Landau levels, we note that the origin $\ket{\mathbf{0}}$ does not correspond to the closest state to the origin $\ket{\mathbf{0}_{\xi}}$ and so care is needed when defining the projector. Specifically, in these cases we define the projector $P_\rho= P_{\rho, \text{LLL}}$, where the LLL subscript denotes that we are using the LLL origin $\ket{\mathbf{0}_{\text{LLL}}}$ in the projector definition, which coincides with the closest state to the origin $\ket{\mathbf{0}_{\xi}}$ for all Landau levels. In this way, we can define $\xi^2$ as the minimum eigenvalue of $P_{\rho,\text{LLL}} D_{n\text{LL}}^2 P_{\rho,\text{LLL}}$, or via $\braket{\mathbf{0}_{n\text{LL}}|P_{\rho,\text{LLL}} D_{n\text{LL}}^2 P_{\rho,\text{LLL}}|\mathbf{0}_{n\text{LL}}}$. Note that if we use $P_{\rho, n\text{LL}}$ for defining the localization length then the scaling exponent $\nu$ is still recovered, however only asymptotically in the large system-size and $\rho\to0$ limits, and so it is more difficult to extract.

\section{Finite-size scaling}
\label{sec:finite_size}

Experimental and numerical studies of critical phenomena are subject to finite-size effects. As one approaches a critical point in the thermodynamic limit, the universal power-law divergences of certain quantities, such as the localization length, is expected. However, in studies of finite systems these divergences are modified as the localization length may not exceed the system size. Using data from multiple finite system sizes to extract the universal behavior in the thermodynamic limit is known as the theory of finite-size scaling. 

The theory of finite-size scaling was introduced in the context of critical phenomena within films of finite thickness by Fisher and Barber in Ref.~\onlinecite{fisher72}. For a more developed perspective, including the field theoretic justification for the finite-size scaling ansatz, we refer the interested reader to the book by Privman~\cite{privman90}. In this section, we review the motivation, techniques, and aspects of the theory~\cite{newman99} that we use in Sec.~III of the main text.

\subsection{Finite-size scaling ansatz}
\label{subsec:finite_size_scaling}

We consider a system parameterized by some dimensionless parameter $\rho$, which undergoes a critical transition at $\rho_c = 0$. Critical phenomena are characterized by a localization length $\xi$, which diverges as a power law with exponent $\nu$, such that $\xi_\infty(\rho) \propto |\rho|^{-\nu}$, in the thermodynamic limit, where the size of the system $L \rightarrow \infty$. Various other quantities, depending on the system definition, may also exhibit singular behavior near the critical point in the thermodynamic limit. As an example, we take a quantity $A$, which diverges in the thermodynamic limit with some exponent $\gamma$, so that $A_\infty(\rho) \propto |\rho|^{-\gamma}$. By rearranging these two power laws, we can write the divergence of $A_\infty$ in terms of $\xi_\infty$ as $A_\infty \propto \xi_\infty^{\gamma/\nu}$.

Given the limitations of experimental and numerical data, we are not able to access infinite systems and therefore the above power-law divergences are not exactly realized. In order to extrapolate data from finite-size systems to infer critical exponents of infinite systems, we review the finite-size scaling ansatz~\cite{newman99}. Consider the data obtained for some $\rho$ in a finite system of linear dimension $L$. If $\xi_\infty(\rho) \ll L$, then we expect $A_L(\rho) \simeq A_\infty(\rho)$. On the other hand, if $L \ll \xi_\infty(\rho)$, we expect the localization length to get cut off at the system size $\xi_L(\rho) \simeq L$, and similarly $A_L(\rho) \simeq L^{\gamma/\nu}$. These considerations motivate the finite-size scaling ansatz for the scaling of the singular quantity $A$ as 
\begin{equation}
\label{eq:fssa}
A_L = \xi_\infty^{\gamma/\nu} f(L/\xi_\infty),
\end{equation}
where $f$ is some dimensionless scaling function, which satisfies
\begin{equation}
f(L/\xi_\infty) \propto \begin{cases} 
\textrm{constant}, & L \gg \xi_\infty \\
\left(L/\xi_\infty\right)^{\gamma/\nu}, & \xi_\infty \gg L
\end{cases},
\end{equation}
and smoothly connects these regimes in the intermediate region where $L \sim \xi_\infty(\rho)$. More commonly used is the rescaled function $\Tilde{f}(x) = x^{-\gamma} f(x^\nu)$, so that Eq.~\eqref{eq:fssa} may be rewritten explicitly in terms of the parameter $\rho$, such that
\begin{equation}
\label{eq:fssa2}
A_L(\rho) = L^{\gamma/\nu}\Tilde{f}(L^{1/\nu}\rho).
\end{equation}

\subsection{Recovery of critical exponents via data collapse}
\label{subsec:data_collapse}

Given data for a singular quantity $A_L(\rho)$ at various $L$ and $\rho$, we now discuss how we can use the finite-size scaling ansatz, in the form of Eq.~\eqref{eq:fssa2}, to recover the critical exponents $\gamma$ and $\nu$. We scale our parameters
\begin{equation}
\label{eq:collapse_a}
\rho \rightarrow \tilde{\rho} = L^{1/\nu} \rho
\end{equation} 
and measurements
\begin{equation}
\label{eq:collapse_b}
A_L(\rho) \rightarrow \tilde{A}_L(\rho) = L^{-\gamma/\nu} A_L(\rho).
\end{equation}
Plotting $\tilde{A}$ against $\tilde{\rho}$ should then result in data across all system sizes $L$ collapsing onto the curve $\tilde{f}$. However, for the collapse to be successful, the scaling in Eqs.~\eqref{eq:collapse_a} and~\eqref{eq:collapse_b} must use the critical exponents $\gamma$ and $\nu$. By varying trial exponents and quantifying the degree of data collapse, we can extract the values of the scaling exponents $\nu$ and $\gamma$. 

In this paper, we use a measure of the quality of data collapse developed in Ref.~\onlinecite{bhattacharjee01}, which we review here. We work with a set of measurements $A_{L_j}(\rho_{i_j})$, where $j \in \{1, 2, \ldots, J\}$ indexes a system size $L_j \in \{L_1, L_2, \ldots, L_J\}$ and $i_j \in \{0, 1, 2, \ldots, I_j\}$ indexes the parameter $\rho$ of a particular measurement at system size $L_j$. For each system size $L_j$, we numerically construct a function $g_j(L^{1/\nu}\rho)$ which interpolates between the scaled measurements $L^{-\gamma/\nu} A_{L_j}(\rho_{i_j})$ with a domain bounded by the set of measurements at $L_j$, that is
\begin{equation}
\label{eq:domain}
L_j^{1/\nu}\min(\{\rho_{i_j}\})\le L^{1/\nu} \rho \le L_j^{1/\nu}\max(\{\rho_{i_j}\}).
\end{equation}
The quality metric $d(\nu,\gamma)$ is then defined as
\begin{equation}
\label{eq:quality}
d(\nu,\gamma) = \left[ \frac{1}{\mathcal{N}} \sum_{j \neq k} \sum_{i_k,\textrm{over}} |L_k^{-\gamma/\nu} A_{L_k}(\rho_{i_k}) - g_j(L_k^{1/\nu} \rho_{i_k})|^q \right]^{1/q},
\end{equation}
where the first sum is over pairs of distinct system sizes indexed by $j$ and $k$, the second sum is over $i_k$ for $\rho_{i_k}$ in the domain of $g_j$ defined in Eq.~\eqref{eq:domain}, $\mathcal{N}$ is the total number of terms in the sum, and $q$ is an integer, which we take to be $q = 2$. The quality function of collapse $d(\nu,\gamma)$ measures the sum of mutual residuals between the scaled data at any two distinct system sizes. Since the scaled parameters $L^{1/\nu} \rho$ for different system sizes need not align, the interpolation functions $g_j$ allow us to approximate the residual.

By minimizing Eq.~\eqref{eq:quality} with respect to $\nu$ and $\gamma$, we will attain the scaled data with the minimal mutual residuals, i.e.~the best data collapse. We use the Nelder-Mead algorithm to minimize $d$~\cite{nelder65}. Using any optimization method, one should be careful to scan a large region of $\nu$-$\gamma$ space to ensure the minimum one finds is not a local minimum. Once the critical exponents $\nu$ and $\gamma$ have been estimated through minimization, we determine the uncertainties $\Delta\nu$ and $\Delta\gamma$ by evaluating the inverse of the Hessian $\partial^2 d(\nu,\gamma)$ at the minimum. 

We note that other measures of the quality of data collapse have been developed in Refs.~\onlinecite{kawashima93},~\onlinecite{houdayer04},~and~\onlinecite{wenzel08}, but are qualitatively equivalent and all correspond to estimating mutual residuals between sets of scaled data.

\section{Statistical moments method for computing the localization length}
\label{sec:moment}

In Secs.~III~A~3 and~III~B~3, we compare the characteristic localization lengths obtained using the minimum eigenvalue of the distance-squared matrix $D^2 = P_{\rho} \mathbf{r}^2 P_{\rho}$, with those obtained using statistical moments of the real-space projector $P(\mathbf{r}, \mathbf{0}) = \braket{\mathbf{r}|P_{\rho}|\mathbf{0}}$. In both cases, we recover the same scaling exponent $\nu\simeq0.5$, as illustrated in Fig.~3 and Figs.~7(a,b), and in the trivial regime, the localization length does not diverge as $\rho\to 0$, as shown in Fig.~7(c). However, despite the close connection between these two metrics, they do not yield identical values of $\xi$. In this section, we explain how the statistical moments are computed, in order to shed light on this discrepancy.

The localization lengths computed using the minimum eigenvalue of the distance-squared matrix are precisely defined via the eigenbasis of $D^2 = P_{\rho} \mathbf{r}^2 P_{\rho}$, and they yield a linear scaling regime with a significantly larger correlation coefficient, compared to the projector moments method, where $\xi^{\gamma} = \braket{r^{\gamma}}$. This can be seen visually by comparing the left panels of Fig.~6 with the right panels of Fig.~7, for example. On the other hand, the statistical moments of the projector yield a more approximate estimate of the localization length, since they are computed directly from the real-space projector plots, shown in the left panels of Figs.~3 and~7. 

The real-space projector is plotted with a certain spatial resolution, and hence we can denote the projector as a discrete function $f(x_i)$, where $\{x\}$ is the set of positions at which $f$ is evaluated. In this notation, the localization length from the $\gamma$th moment of the projector is
\begin{equation}
\xi = \left[ \sum_{i} p(x_i) x_i^{\gamma} - \left( \sum_i p(x_i) x_i \right)^{\gamma} \right]^{\frac{1}{\gamma}},
\end{equation}  
where the probability $p(x_i) = f(x_i) / \sum_i f(x_i)$. In Figs.~3 and 7, the spatial resolution of the projectors is $\Delta r = 0.5$ and so it is clear that appreciable sampling error is introduced by this method. Moreover, due to numerical effects, the plotted projectors are not always precisely symmetric about $r=0$, which may lead to a non-zero mean $\sum_i p(x_i)x_i \neq 0$ that can affect the estimate of $\xi$. 

Nevertheless, despite inherent differences between the two approaches, as well as the crude method of extracting the moments, the values of $\xi$ generally agree within $\sim 20\%$ for Landau levels and $\sim 40\%$ for the Haldane model. The values extracted for Landau levels in Fig.~3 show closer agreement than those extracted for the Haldane model in Fig.~7, since the Haldane model additionally suffers from lattice effects. The trivial regime comparison between Fig.~6(c) and Fig.~7(c) shows a notably larger discrepancy, where the moments method yields a $\lim_{\rho\to 0}\xi$ value $\sim 2.5$ times larger than the distance-squared matrix method. In this case, the unusually large difference stems from the irregular form of the projector under this renormalization procedure, which exacerbates numerical errors.

\section{Parameter dependence of localization scaling in Chern insulators}
\label{sec:param}

In Sec.~III~B, we study the localization scaling for the Haldane model at three different points in the phase diagram, $\{\mathbf{t}_a, \mathbf{t}_b, \mathbf{t}_c\}$, depicted in Fig.~4(b). Our motivation for selecting these parameters sets is twofold. First, we choose points that are diverse, i.e.~we pick two points in the topological phase ($\mathbf{t}_a$ with $M=0$ and $\mathbf{t}_b$ with $M\neq0$, such that $t_{2,a}\neq t_{2,b}$), and one point in the trivial phase ($\mathbf{t}_c$). With this selection, we can verify that the localization scaling holds in the topological phase at different values of $(t_2, M)$, and also contrast this with behavior in the trivial regime. Second, we choose points deep in their respective phases, such that we can obtain an accurate estimate for the scaling exponent $\nu$ at low numerical cost. As we move towards a topological phase transition, the required system size to recover $\nu\simeq0.5$ with high precision increases, and so we select parameters that facilitate an efficient convergence of $\nu$, using system sizes comparable to Sec.~III~A.

\begin{figure}
	\centering
	\includegraphics[width=.5\linewidth]{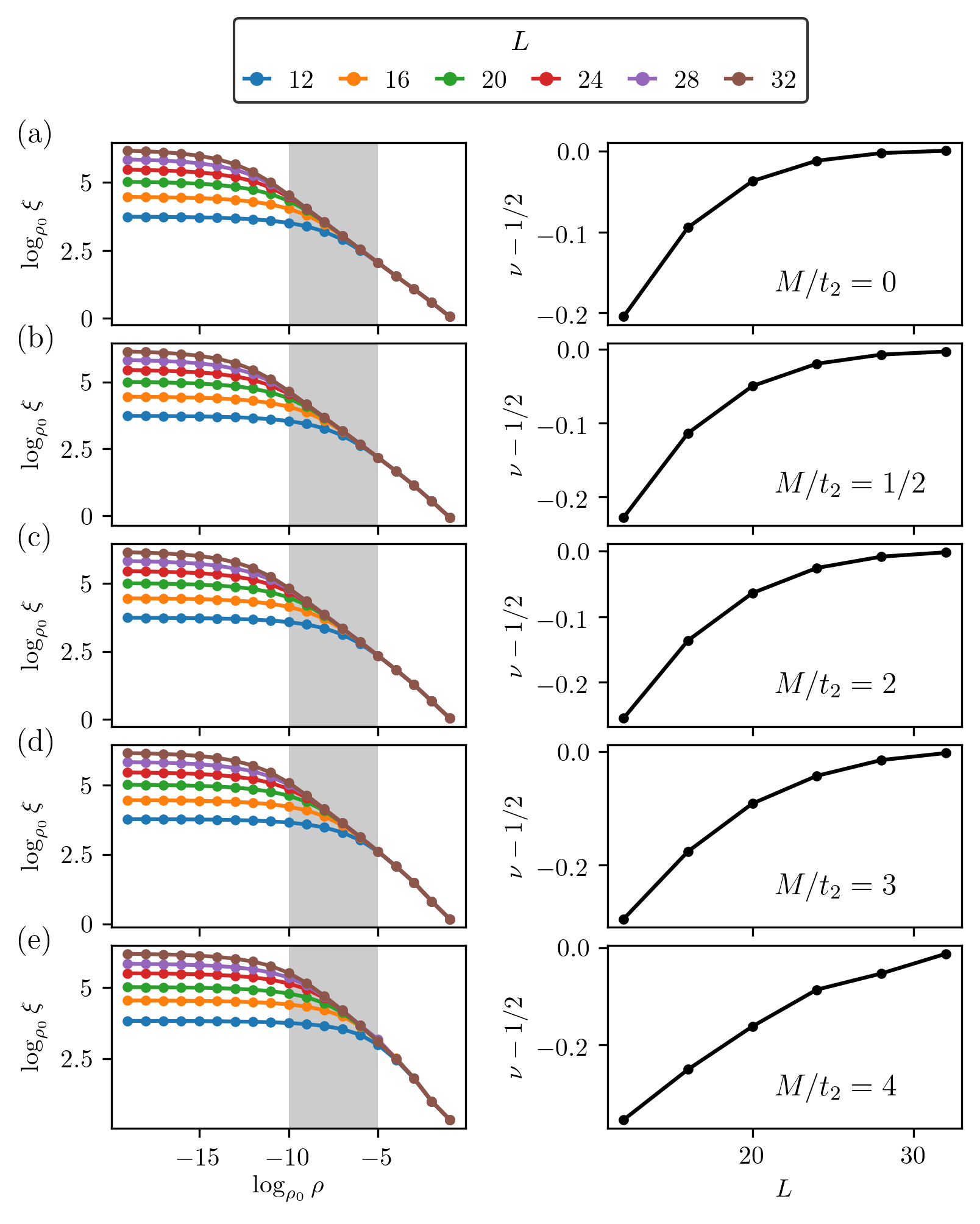}
	\caption{\label{fig:param} \textbf{Scaling for different parameter values in the Haldane model.} (left)~Localization length $\xi(\rho)$ obtained for the Haldane model, using the minimum eigenvalues of the distance-squared matrix, as described in Sec.~III~B~2. The scaling is shown for a variety of parameter sets, with $t_1=1$ and $\phi=\pi/2$ corresponding to Fig.~4(b), and $\rho_0=1.5$. The parameters are chosen such that we have five distinct values of $M/t_2$ in the topological phase, which approach the boundary at $M/t_2=3\sqrt{3}$. (a) corresponds to $\mathbf{t}_a$ with $\{t_2=0.1, M=0\}$, (b) corresponds to $\mathbf{t}_b$ with $\{t_2=0.2, M=0.1\}$, and the (c--e) maintain $t_2=0.2$ while increasing $M$. (right)~Scaling exponent $\nu$ from $\lim_{\rho\to 0}\xi(\rho)\sim\rho^{-\nu}$, approximated using lines of best fit in the left plots. The lines of best fit are constructed using the $\rho$ domain corresponding to the six smallest values in the linear regions of the $L=32$ curves in (a,b), which is shaded gray.}
\end{figure}

In Fig.~\ref{fig:param}, we show the localization scaling at five distinct values of $M/t_2=0, 1/2, 2, 3, 4$, in the topological phase, which gradually approach the boundary at $M/t_2=3\sqrt{3}$. For comparison with the main text, we set $t_1=1$ and $\phi=\pi/2$ throughout, and compare the systems sizes $L=12, 16, 24, 28, 32$. Figures~\ref{fig:param}(a,b) correspond to the parameter sets $\mathbf{t}_a$ and $\mathbf{t}_b$, whereas Figs.~\ref{fig:param}(c--e) maintain $t_2=0.2$, in correspondence with $\mathbf{t}_b$. In contrast to the main text, however, we do not obtain $\nu$ from a finite-size scaling collapse, since this leverages all system-size data. Instead, we approximate $\nu$ from the gradients of trend lines in our left plots, for each $L$ individually. Moreover, since the scaling exponent is only recovered asymptotically in the $\rho\to 0$ limit, we consider the $\rho$ domain corresponding to the six smallest values in the $L_{\text{max}}$ linear regions of (a,b) to construct our lines of best fit. Although this method does not yield precise estimates for $\nu$, it does show how $\nu$ converges relative to $L$, for each $M/t_2$.

From Fig.~\ref{fig:param}, we can see that $\nu$ converges most quickly for parameters deep in the topological phase, $\mathbf{t}_a$ and $\mathbf{t}_b$, which supports our choice in the main text. In these cases, we can recover an accurate estimate of $\nu=0.5$ using only our chosen range of $L$. As we increase the value of $M/t_2$, however, this convergence becomes more expensive. In the right panels of Figs.~\ref{fig:param}(c--e), we can see that the slope becomes mellower and therefore the convergence to $\nu\simeq0.5$ is slower. Moreover, the estimates of $\nu$ from small system-size data become significantly worse, where for the $L=12$ curves, the absolute deviation from $\nu=0.5$ grows monotonically as $M/t_2$ increases. Extrapolating the $\nu$-$L$ curves for higher $M/t_2$ values shows that we can recover the same high-precision scaling exponent at larger system sizes, as expected. However, reaching the system sizes needed to verify the scaling exponent for parameter sets close to a topological phase transition, is numerically inaccessible, using our current approach.  

\bibliographystyle{apsrev4-1}
\bibliography{ms}